\documentclass[useAMS,usenatbib,usegraphicx]{mn2e}
\usepackage{psfig}   
\usepackage{graphicx}
\usepackage{subfigure}

\newcommand{\apj}{ApJ}                                         
\newcommand{\apjs}{ApJS}
\newcommand{\aap}{A{\&}A}

\newcommand{\mnras}{MNRAS}
\newcommand{\aj}{AJ}
\newcommand{\apss}{ApSS}

\newcommand{\araa}{ARA\&A}

\newcommand{\Msun}{\ensuremath{\textrm{ M}_{\odot}}}
\newcommand{\kms}{\ensuremath{\textrm{ km s}^{-1}}}
\newcommand{\nth}{\ensuremath{^{\rm th}}}

\title[The early dynamical evolution of cool, clumpy star clusters]{The early dynamical evolution of cool, clumpy star clusters}

\author[R.~J.~Allison et al.]{Richard~J.~Allison$^1$, Simon
  P.~Goodwin$^1$\thanks{E-mail: s.goodwin@sheffield.ac.uk},
  Richard~J.~Parker$^1$,\newauthor Simon~F.~Portegies~Zwart$^2$, and
  Richard~de~Grijs$^{3,1}$
  \vspace*{0.1cm}\\ $^1$ Department of Physics and Astronomy,
  University of Sheffield, Sheffield, S3 7RH, UK \\ $^2$ Leiden
  Observatory, Leiden University, PO Box 9513, 2300 RA Leiden, The
  Netherlands \\ $^3$ Kavli Institute for Astronomical Astrophysics,
  Peking University, Beijing 100871, China}

\begin{document}

\date{}
                             
\pagerange{\pageref{firstpage}--\pageref{lastpage}} \pubyear{2010}

\maketitle

\label{firstpage}

\begin{abstract}
Observations and theory both suggest that star clusters form
sub-virial (cool) with highly sub-structured distributions.  We
perform a large ensemble of $N$-body simulations of moderate-sized
($N=1000$) cool, fractal clusters to investigate their early dynamical
evolution.  We find that cool, clumpy clusters dynamically mass
segregate on a short timescale, that Trapezium-like massive
higher-order multiples are commonly formed, and that massive stars are
often ejected from clusters with velocities $> 10$~km~s$^{-1}$
(c.f. the average escape velocity of 2.5~km~s$^{-1}$).  The
properties of clusters also change rapidly on very short timescales.
Young clusters may also undergo core collapse events, in which a
dense core containing massive stars is hardened due to energy losses
to a halo of lower-mass stars.  Such events can blow young clusters
apart with no need for gas expulsion.  The warmer and less
substructured a cluster is initially, the less extreme its evolution.
\end{abstract}

\begin{keywords}   
methods: $N$-body simulations - stars: formation - stars: kinematics
and dynamics
\end{keywords}

\section{Introduction}
\label{sec:intro}

Most stars appear to form in star clusters
\citep{lada03,lada09,portegies_zwart10} and so star formation is
inextricably linked with star cluster formation.  Recent advances in
observations and theory have allowed us to construct a basic picture
of cluster formation in which clusters form dynamically cool
(sub-virial), and highly substructured.

Clusters form in highly turbulent molecular clouds.  These clouds are
highly substructured, containing dense clumps and filaments
\citep{williams99,williams00,carpenter08} which are presumably formed
by the decay of supersonic turbulence
\citep{maclow04,ballesteros-paredes07}. Stars and small stellar groups
form in these dense regions and are unsurprisingly observed to have a
high degree of substructure when young
\citep{larson95,elmegreen00,testi00,cartwright04,gutermuth05,allen07,schmeja06,schmeja08},
clustering in young stellar groups has even been observed in the SMC
\citep{schmeja09} and LMC \citep{bastian09}.  The same behaviour is
seen in simulations of cluster formation
\citep{klessen00,klessen01,bate03,bonnell03,bate09,offner09},
including in comparison tests between AMR and SPH techniques
\citep{federrath10}.

Clusters are observed to lose their substructure as they evolve,
becoming smooth and roughly spherical
\citep{cartwright04,schmeja06,schmeja08}. \citet{goodwin04} have shown
that substructure can only be erased in clusters if the clusters are
initially cool (sub-virial) (see also Maschberger et al. 2010).  
Both observations of pre-stellar cores
\citep{belloche01,andre02,walsh04,peretto06,kirk07} and stars
\citep{peretto06,proszkow09} show that they indeed appear to be
sub-virial, a property also found in simulations of cluster formation
(Klessen \& Burkert 2000; Offner et al. 2009; Maschberger et al. 2010).

Following Allison et al. (2009b) we conduct $N$-body simulations of a
large number of initially sub-virial, fractal star clusters.  In
Section~\ref{sec:init} we describe our simulations. In
Section~\ref{sec:results} we describe our main results, specifically
the early onset of dynamical mass segregation and interesting
`post-collapse' evolution. In Section~\ref{sec:disc} we discuss the
results, and we summarise and conclude in Section~\ref{sec:conc}.

\section{Initial Conditions}
\label{sec:init}

We perform 160 $N$-body simulations with 1000 stars each, in which the
initial conditions are cool and clumpy.  We vary the level of
substructure and initial virial ratio, and conduct ensembles of
simulations with the same initial conditions, varying only the initial
random number seed used to initialise the simulations.

To create initial substructure in our simulations we use a fractal
stellar distribution. Using a fractal distribution provides a
parameterisation of substructure using only a single number: the
fractal dimension.  (Note that we are not claiming that clusters are
actually initially fractal, although they may be
\citep{elmegreen01,cartwright04}, just that this provides a simple
descriptor of substructure that is easy to reproduce).

The fractal stellar distributions were generated following the
method of \citet{goodwin04}. The method begins by defining a cube
of side $N_{{\rm div}}$ (we use $N_{{\rm div}} = 2$ throughout), 
inside of which the fractal will be built. A
first-generation parent is placed at the centre of the cube, from
which are spawned $N_{{\rm div}}^{3}$ sub-cubes, each containing a
first-generation child in its centre. The fractal is then built
by determining which of the children themselves become parents,
and spawn their own offspring. This is determined by the fractal
dimension, $D$, where the probability that a child becomes a
parent is $N_{\rm div}^{(D-3)}$. For a lower fractal dimension
less children will mature and so the final distribution will
contain more structure. Any children which do not become parents
in a given step are removed, along with all of their parents. A
small amount of noise is then added to the positions of the
remaining children, preventing the final cluster from having
a gridded appearance, and the children become parents of the next
generation. Each new parent then spawns $N_{\rm div}^{3}$
second-generation children in $N_{\rm div}^{3}$ sub-sub-cubes,
with each second-generation child having a $N_{\rm div}^{(D-3)}$
probability of becoming a second-generation parent. This process
is then repeated until there are substantially more children than
required. The children are pruned to produce a sphere from the
cube and are then randomly removed (so maintaining the fractal
dimension) until the required number of children are left. These
children then become the stars in the cluster.

To determine the velocity structure of the cloud, children inherit
their parent's velocity plus a random component that decreases with
each generation in the fractal.  The children of the first generation
are given random velocities from a Gaussian of mean zero\footnote{Here
  the variance is unimportant as the velocities are scaled to the
  desired virial ratio once the final spatial distribution is
  obtained, but the method can also be used to match the
  \citet{larson81} relations if desired
  \citep[see][]{goodwin04}.}. Each new generation then inherits their
parent's velocity plus an extra random component that becomes smaller
with each generation.  This results in a velocity structure in which
nearby stars have similar velocities, but distant stars can have very
different velocities.  Finally, the velocity of every star is scaled
to obtain the desired total virial ratio for the cluster.

%----tab:runs----------
\begin{center}
\begin{table}
  \begin{tabular}{|c|c|c|c|c|}
\hline
        &  \multicolumn{4}{|c|}{$D$} \\
$Q$     &  1.6        &  2.0        &  2.6        &  3.0  \\
\hline
0.3     &  a1.01--50  &  a2.01--10  &  a3.01--10  &  a4.01--10   \\
0.4     &  b1.01--10  &  b2.01--10  &  b3.01--10  &  b4.01--10   \\
0.5     &  c1.01--10  &  c2.01--10  &  c3.01--10  &  c4.01--10   \\
\hline
  \end{tabular}
\caption{Notation for run identification where $D$ is the initial
  fractal dimension, and $Q$ is the initial virial ratio of each
  simulation. The numbers 01--50 and 01--10 are the identifiers for
  each individual run.  Within each ensemble only the random number
  seed used to generate the initial conditions is
  changed. \label{tab:runs}}
\end{table}
\end{center}
%----tab:runs----------

The simulations contain 1000 stars, have an initial maximum radius of
1 pc, include no primordial binaries or gas and a three-part power law
is used to produce an initial mass function \citep[IMF,][]{kroupa02},

\begin{equation}
  N(M) \propto \left\{
  \begin{array}{r}
    M^{-0.3} \quad m_0 \leq M/\Msun < m_1, \\
    M^{-1.3} \quad m_1 \leq M/\Msun < m_2, \\
    M^{-2.3} \quad m_2 \leq M/\Msun < m_3, \\ 
  \end{array}
  \right.
\end{equation}

\noindent with $m_0=0.08\Msun$, $m_1=0.1\Msun$, $m_2=0.5 \Msun$ and
$m_3=50\Msun$. No stellar evolution is included because of the short
duration of the simulations ($\sim 4$ Myr).  We use the {\sc starlab}
$N$-body integrator {\sc kira} to run our simulations
\citep{portegies_zwart01}.

In this study we explore a range of fractal dimensions and virial
ratios. The fractal dimensions investigated are
$D=1.6,~2.0,~2.6~\textrm{and}~3.0$ (since these values correspond to
the number of maturing children, $N_{{\rm div}}^D \equiv 2^{D}$ ,
being an integer), where $D=1.6$ produces a large amount of structure,
and $D=3.0$ produces a uniform sphere. We investigate virial ratios of
$Q = 0.3,~0.4~\textrm{and}~0.5$, we define the virial ratio as
$Q=T/|\Omega|$ (where $T$ and $|\Omega|$ are the total kinetic and
total potential energy of the stars, respectively), hence virial
equilibrium is $Q=0.5$.

It is important to note that fractal initial conditions are inherently
stochastic: statistically identical fractals (i.e., the same fractal
dimension), can appear very different to the eye, and can evolve in
very different ways (see Section~\ref{ssec:stocasticity}).  Therefore,
it is vital to perform large ensembles of simulations with different
random number seeds.  We have therefore simulated 50 $D=1.6$, $Q=0.3$
(a1) clusters (as they have the most interesting evolution), and
restricted our analysis of all other combinations of $D$ and $Q$ to 10
clusters each.  The identifiers and initial conditions of each 
ensemble are presented in table~\ref{tab:runs}.

To quantify mass segregation we use the minimum spanning tree (MST)
method \citep{allison09}. This method compares the MST of the $N$ most
massive stars with the average MSTs of $N$ randomly selected stars.
The ratio of these two MST lengths gives a quantitative measure of the
concentration of the massive stars and hence a value of the mass
segregation in the cluster,

\begin{equation}
\Lambda=\frac{\langle l_{\rm norm} \rangle}{l_{\rm massive}}\pm
\frac{\sigma_{\rm norm}}{l_{\rm massive}}
\end{equation}

\noindent where $\Lambda$ is the measure of mass segregation,
${\langle l_{\rm norm} \rangle}$ is the average length of the random
MSTs, $l_{\rm massive}$ is the length of the massive star MST and
$\sigma_{\rm norm}$ is the error in the random MST length. The value
$\sigma_{\rm norm}/l_{\rm massive}$ is the 1 $\sigma$ error in
$\Lambda$.

In this paper, mass segregation is calculated for the entire stellar
content (i.e. ejected stars are not removed from the simulation) in
all three dimensions. We note that an observer will see only two
dimensions, and may not identify all cluster members, in particular
low-mass stars and stars that have been ejected from the cluster.
Allison et al. (2009a) show that there is little difference between
the mass segregation measure in two and three dimensions when using
mass segregated Plummer spheres. However, as with all mass segregation
measures, if using the measure in two dimensions projection effects
may cause differences depending on orientation. Unfortunately, this is
unavoidable. Including all stars is useful theoretically, however it
does not allow a direct comparison with observational data.

\section{Results}
\label{sec:results}

\subsection{Rapid dynamical mass segregation}
\label{ssec:dyn_ms}

%----fig:a1.34----------f1000.a1.34; Snapshot #: 001,081,102
%NEED TO IDL THESE IMAGES
\begin{figure*}
  \begin{center}
    \setlength{\subfigcapskip}{10pt}
\subfigure[0 Myr]{\label{fig:a1.34.001}
\includegraphics[scale=0.5,angle=270]{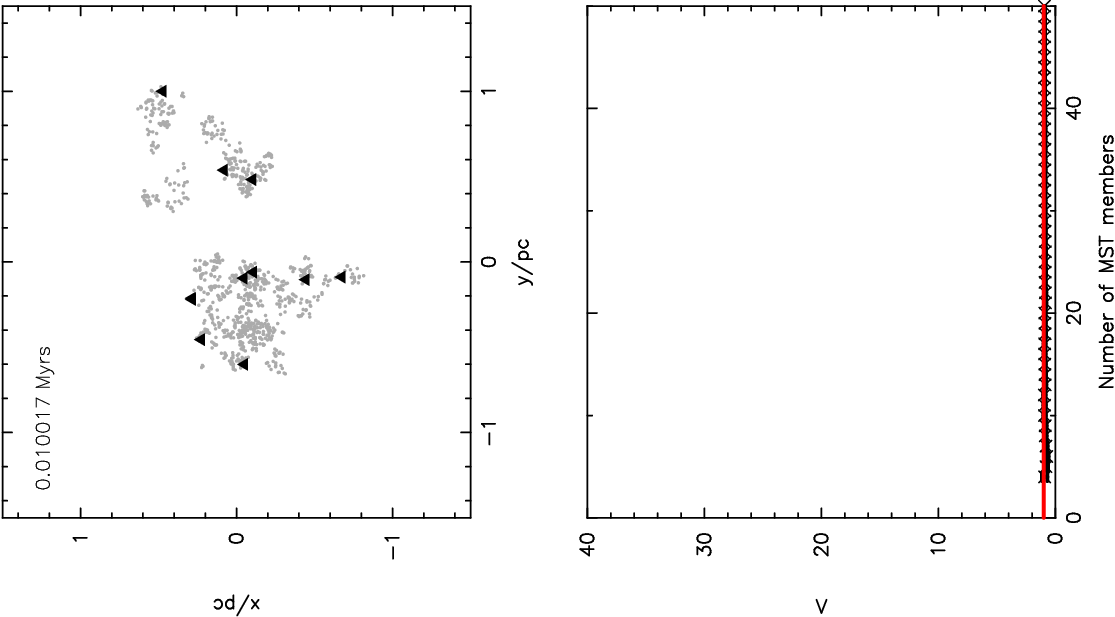}}
\subfigure[0.8 Myr - Formation of core]{\label{fig:a1.34.081}
\includegraphics[scale=0.5,angle=270]{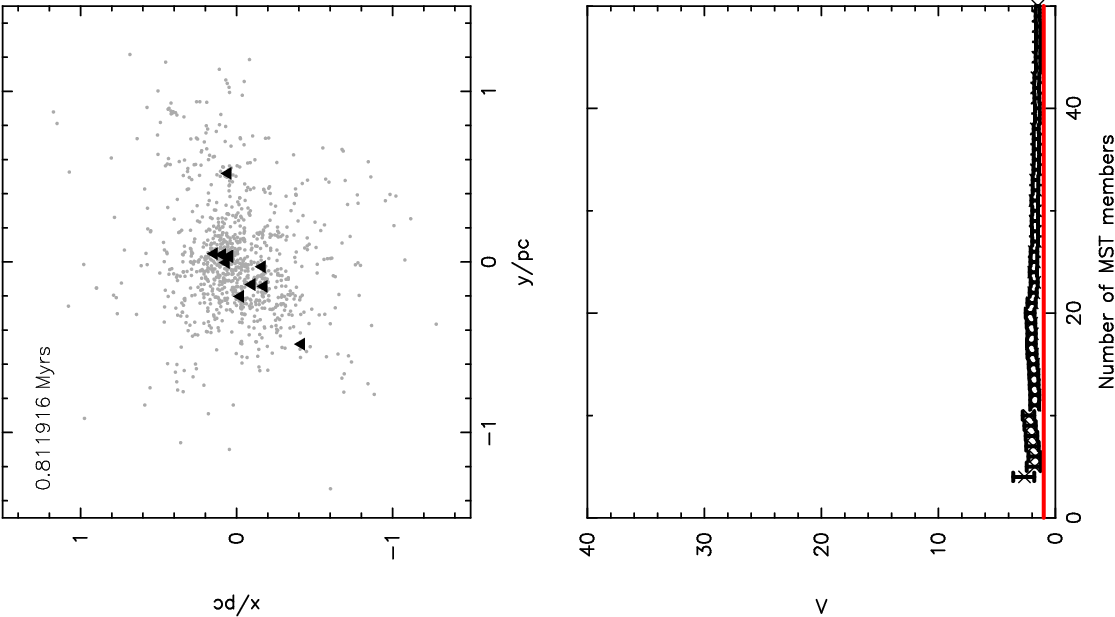}}
\subfigure[1 Myr - End of core]{\label{fig:a1.34.102}
\includegraphics[scale=0.5,angle=270]{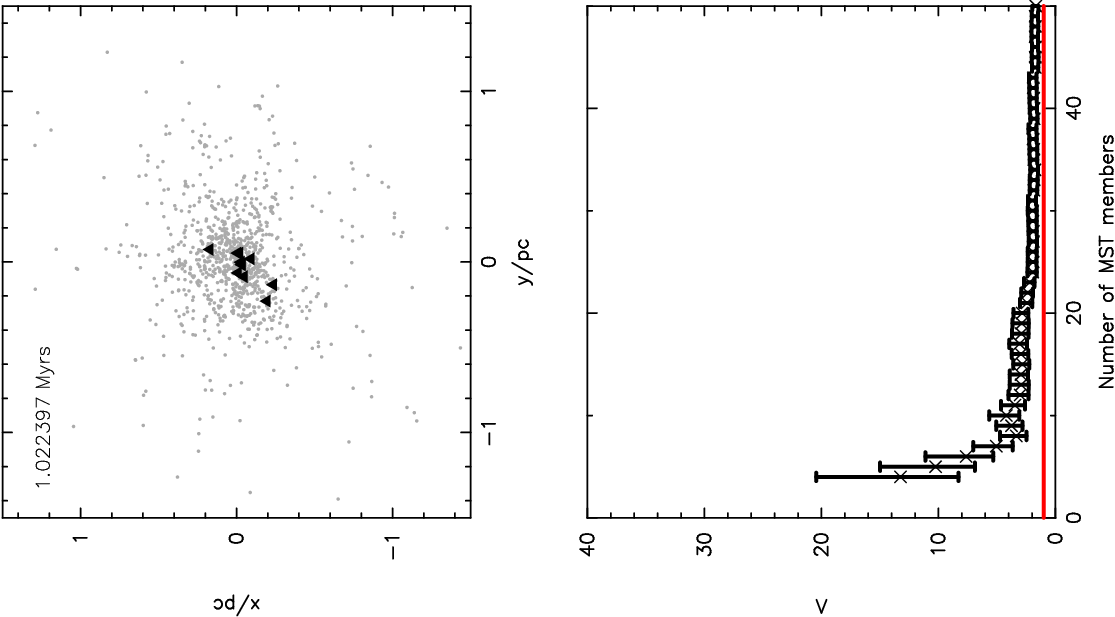}}
%\hspace*{0.6cm}
  \end{center}
  \caption[bf]{Run a1.34 -- \textit{Top:} 2 dimensional stellar
    distributions at (a) 0 Myrs - the initial distribution, (b) 0.8
    Myrs - the formation of the dense core and (c) 1 Myr - the end of
    the dense core and initial dynamical mass segregation. Triangles
    indicate the positions of the 10 most massive
    stars. \textit{Bottom:} The evolution of $\Lambda$ with $N$ for
    (a), (b), (c); as described above. The error bars show $1\sigma$
    deviations.}
  \label{fig:a1.34}
\end{figure*}
%----fig:a1.34----------

In \citet[][hereafter Paper I]{allison09b} we showed that cool
($Q=0.3$), and clumpy ($D=1.6$) clusters can dynamically mass
segregate on a timescale close to the initial crossing time of the
system ($\sim 1$~Myr).  Figure~\ref{fig:a1.34} shows the early
evolution and onset of mass segregation in such a cluster. The top
panels show the spatial distributions of the cluster at times $\sim 0.01,
0.81$ and $1.02$~Myr (left to right), whilst the lower panels show the
mass segregation ratio $\Lambda$ for each of these snapshots -- the
higher the value of $\Lambda$, the greater the degree of mass
segregation.  As is clear from the figure, the cluster evolves from an
initially non-mass segregated and clumpy distribution, into a smooth
and mass segregated one.

In paper I we argued that this rapid dynamical mass segregation is due
to the collapse of the cool cluster forming a very dense, but
short-lived core (at $\sim 0.8$~Myr in the example shown in
Fig.~\ref{fig:a1.34}).  These dense cores (for $D=1.6$, $Q=0.3$) tend
to contain about half of the mass of the cluster, are roughly $0.1$ --
$0.2$~pc in radius, and survive for $0.1$ -- $0.2$~Myr ($10$ -- $20$
crossing times in the core).

\citet{spitzer69} showed that the timescale for mass segregation,
$t_{\rm seg}$, for a star of mass $M$ depends on how massive that star
is relative to the average mass of a star in a cluster, $\langle m
\rangle$,

\begin{equation}
t_{\rm seg }(M) \approx \frac{\langle m \rangle}{M} t_{\rm relax}, 
\label{eq:tseg}
\end{equation}

where $t_{\rm relax}$ is the relaxation time of the cluster, which is
related to its crossing time, $t_{\rm cross}$, by

\begin{equation}
t_{\rm relax }\approx \frac{N}{8\ln N} t_{\rm cross}. 
\label{eq:trelax}
\end{equation}

Eq.~\ref{eq:tseg} can be rewritten as

\begin{equation}
t_{\rm seg } \approx \frac{\langle m \rangle}{M}
\frac{N}{8\ln{N}} t_{\rm cross }. 
\label{eq:tdense.exp}
\end{equation}

For the $D=1.6$, $Q=0.3$ initial conditions used in paper I, typical
values for these parameters are $N \sim 300$ -- $500$, $t_{\rm cross }
\sim 0.01$ -- $0.2$~Myr, $\langle m \rangle \sim 0.4 \Msun$ (typical
for standard IMFs).  The core has a lifetime of $0.1$ -- $0.2$~Myr in
which it can mass segregate giving a mass to which the core can
segregate of $M \sim 2$ -- $4 \Msun$.  In Fig.~\ref{fig:a1.34} the
cluster is mass segregated to the 20th most massive star, which has a
mass of a few $\Msun$, in good agreement with our simple analytical
model.

The reason that a dense core can form is that the cluster is
both cool {\em and} clumpy: cool clusters will initially collapse, but
cool and clumpy clusters can collapse further.  The potential
energy, $\Omega$, of a cluster of mass $M_{\rm clus}$ and radius $R$ is

\begin{equation}
\Omega = - \eta \frac{GM_{\rm clus}^2}{R},
\end{equation}

\noindent where $\eta$ is a structure parameter whose value depends on
the choice of $R$ (e.g., is it the core radius or the half-mass
radius?), and the structure of the cluster (e.g., is it clumpy,
Plummer, or uniform density?) \citep[see also,][]{portegies_zwart10}.
For example, for a Plummer sphere, if $R$ is the Plummer radius then
$\eta \sim 0.3$.

If a cluster has an initial potential energy $\Omega_0$ (with radius 
$R_0$ and structure parameter $\eta_0$), and an initial virial ratio
$Q_0$, then the initial total energy $E_0$ is

\[
E_0 = -\eta_0 \frac{GM_{\rm clus}^2}{R_0} (1 - Q_0).
\]

Whatever the initial conditions, a (bound) cluster will attempt to
reach virial equilibrium and a relaxed configuration (something like a
Plummer sphere, or King profile with a concentration
parameter\footnote{We define the concentration parameter as $c={\rm
    log}_{10}(r_{\rm virial}/r_{\rm core})$} $\approx 2-3$, for an a1
type cluster).  Therefore, the final energy $E_f$ of the cluster (with
potential energy $\Omega_f$, radius $R_f$, structure parameter
$\eta_f$, and virial ratio $Q_f=0.5$) will be

\[
E_f = -\eta_f \frac{1}{2} \frac{GM_{\rm clus}^2}{R_f}
\]

\noindent (assuming no mass is lost).  Equating these equations gives
the degree of collapse (or expansion if initially warm) of the cluster

\begin{equation}
\frac{R_0}{R_f} = \frac{\eta_0}{\eta_f} 2(1 - Q_0).
\end{equation}

Clearly, to induce dynamical mass segregation on a short timescale,
$R_0/R_f$ must be as large as possible to cause the maximum degree
of collapse.  This implies that both $\eta_0/\eta_f$ and $2(1 -
Q_0)$ need to be large.

Obviously, making $2(1 - Q_0)$ large means making $Q_0$ as small as
possible, but even with an initially static stellar distribution with
$Q_0 = 0$, $2(1 - Q_0)$ can never be greater than two (this is why
Bonnell \& Davis 1998 did not see rapid, early mass segregation).
Realistically, it is difficult to imagine $Q_0$ being much less than
$0.2$ or $0.3$ as the stars {\em must} initially have some relative
velocities.

As stated above, the cluster will not just relax into virial
equilibrium, it will also attempt to reach a basic statistical
equilibrium which is a smooth, centrally concentrated distribution 
(like a Plummer sphere or King model if tidally truncated).  Therefore
the final structure parameter will be $\eta_f \sim 0.3$ when the
scale radius is the Plummer radius.  

Numerical experiments we have carried out show that the structure
parameter for a fractal of dimension $D=1.6$ is $\eta_0 \sim 1.1 \pm
0.1$, for $D=2.0$, $\eta_0 \sim 0.8 \pm 0.1$ (note that different
realisations of $D=1.6$ and $D=2.0$ fractals have a large variation in
their structure parameters), for $D=2.6$, $\eta_0 \sim 0.7$, and for
$D=3.0$, $\eta_0 \sim 0.6$.  For each of these $\eta$ values the scale
radius is the total radius of the fractal.  Unfortunately, it is very
difficult to exactly compare the $\eta$ of a Plummer model and the
$\eta$ of a fractal as the radii are defined differently.  The
half-mass radius of fractal models varies quite significantly (and its
meaning is also rather unclear in a fractal) and so does not provide a
useful comparison radius either.

\subsection{Mass segregation in clusters with different $D$ and $Q$}
\label{ssec:DandQ}

The analysis above suggests that the initial virial ratio and the
initial fractal dimension of a cluster are the crucial parameters in
determining whether that cluster will be able to rapidly dynamically
mass segregate.  In particular, the warmer (higher-$Q$), and the
smoother (higher-$D$) a cluster is, the lower the maximum central
density, and hence less dynamical mass segregation will occur.

\begin{figure*}
  \begin{center}
    \setlength{\subfigcapskip}{10pt}
\subfigure[$D=1.6; Q=0.3$ -- a1.21]{\label{fig:a1.21}
\includegraphics[scale=0.25,angle=270]{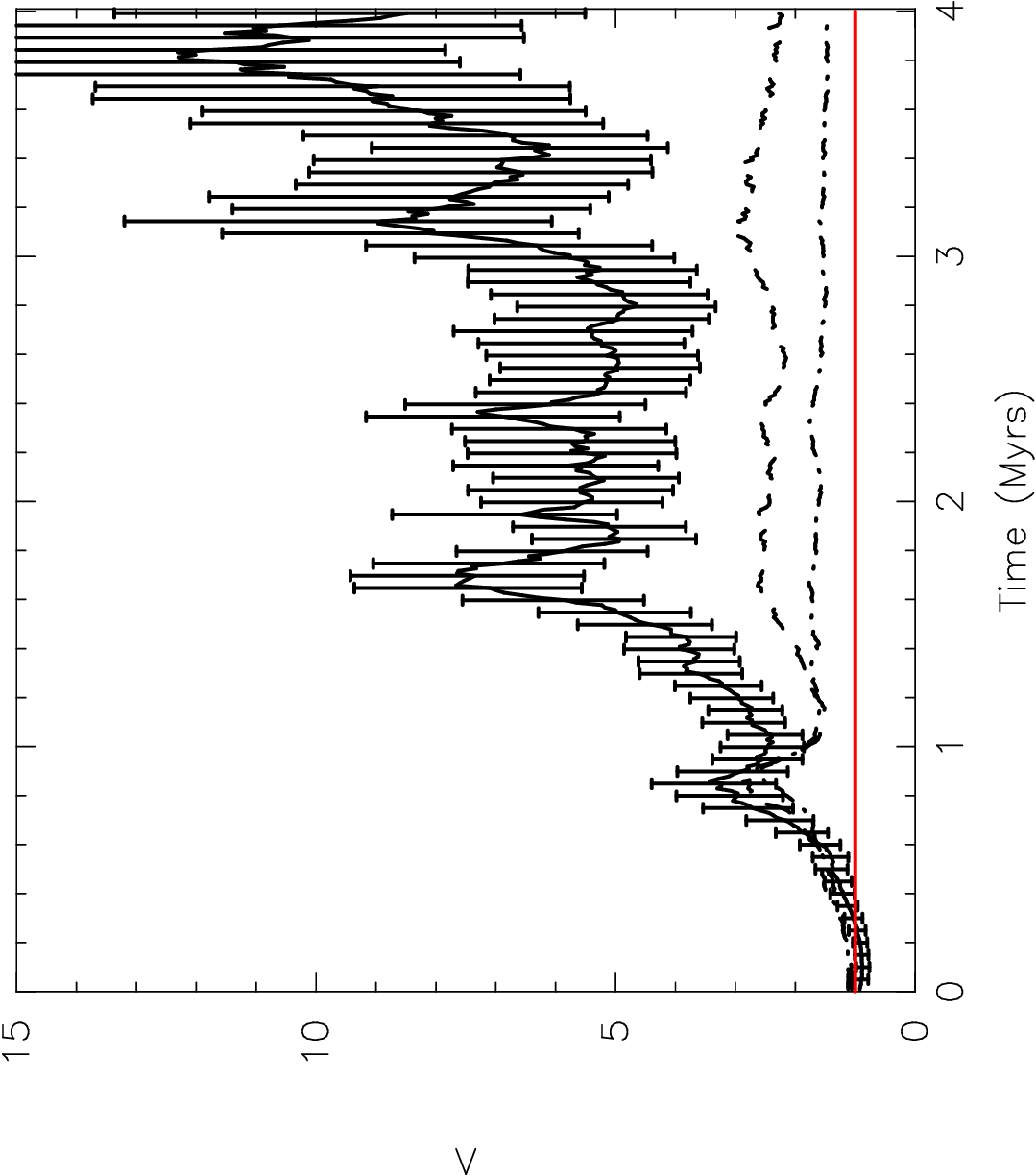}}
\subfigure[$D=1.6; Q=0.4$ -- b1.05]{\label{fig:b1.05}
\includegraphics[scale=0.25,angle=270]{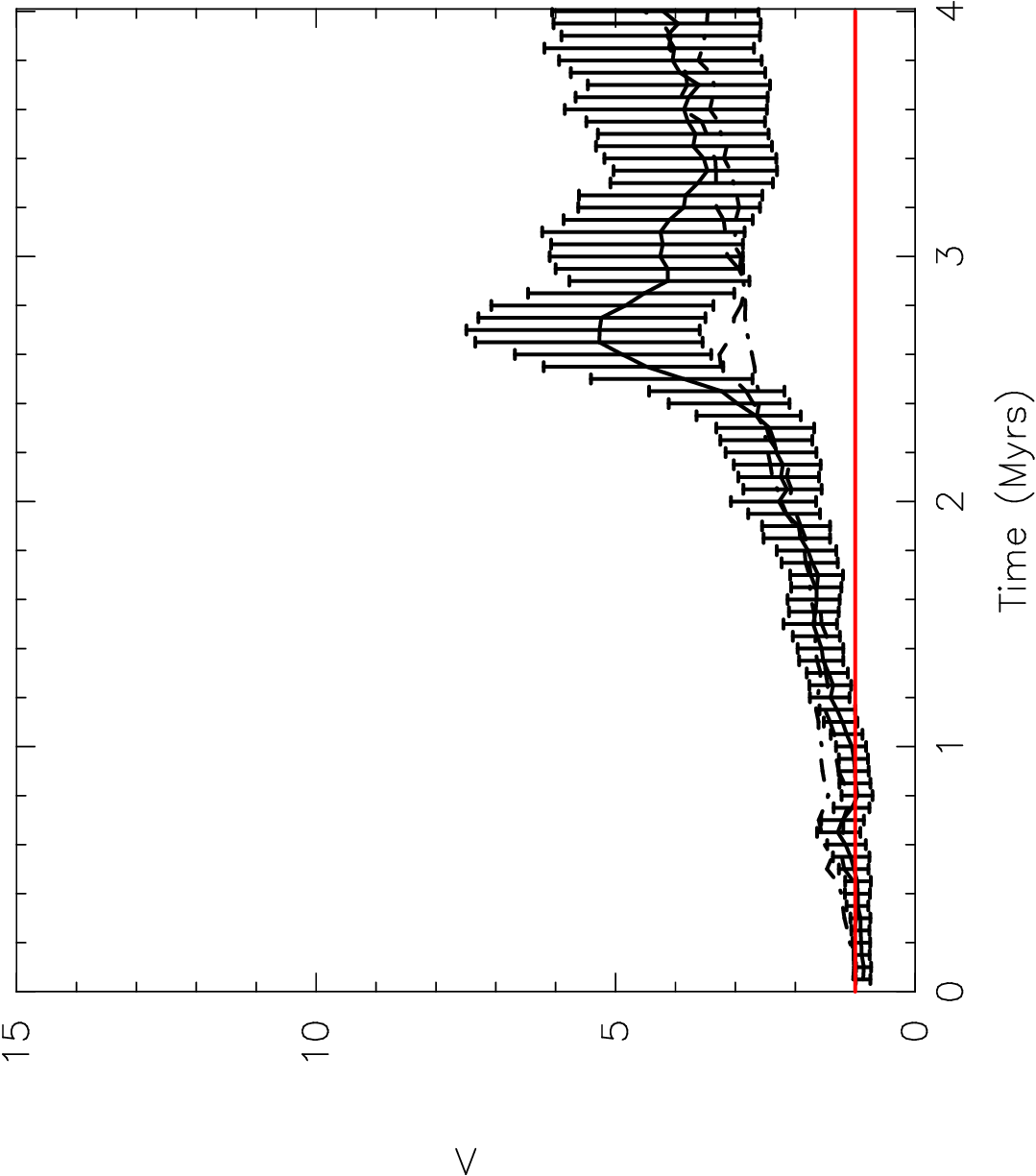}}
\subfigure[$D=1.6; Q=0.5$ -- c1.06]{\label{fig:c1.06}
\includegraphics[scale=0.25,angle=270]{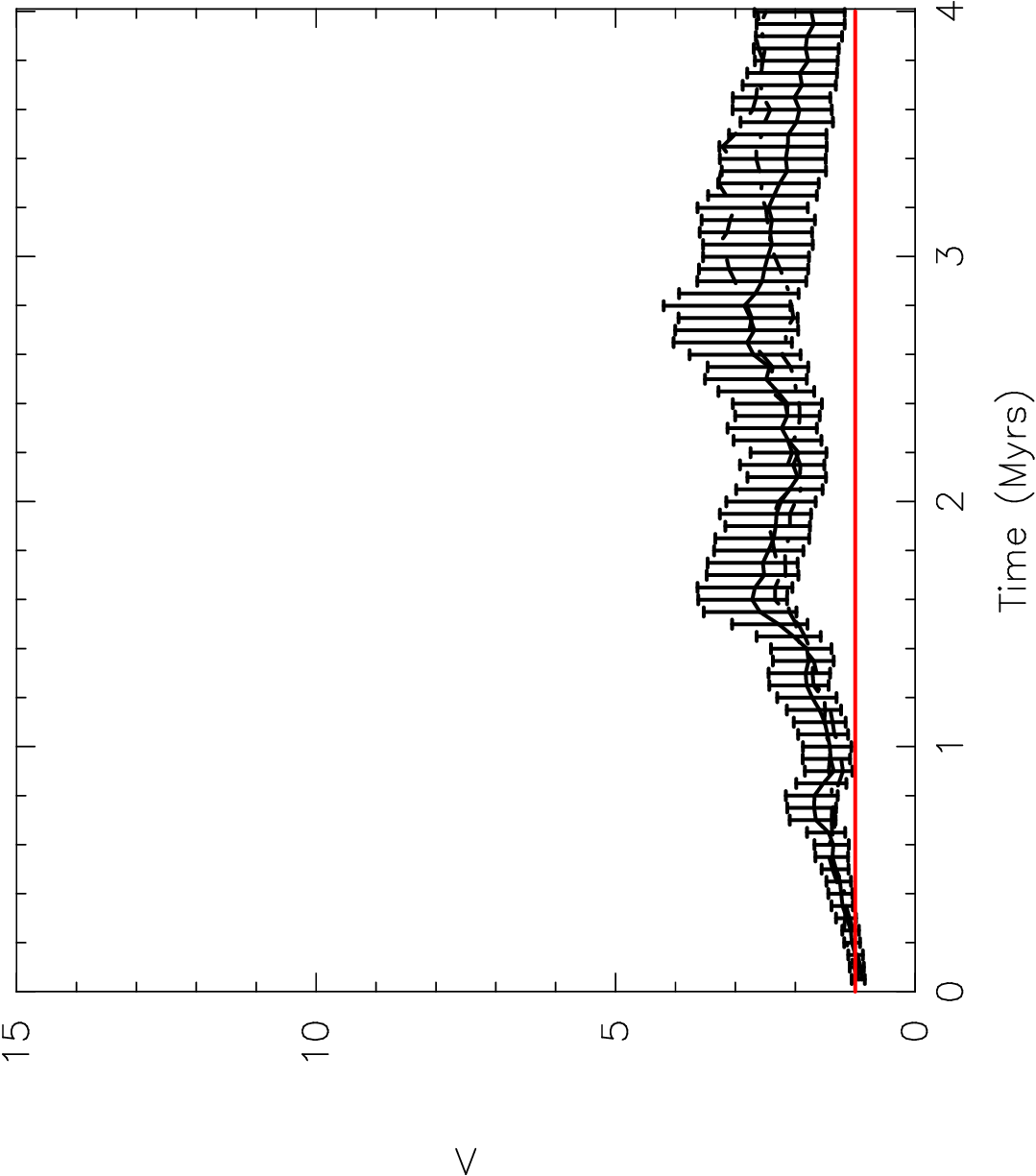}}

\subfigure[$D=2.0; Q=0.3$ -- a2.08]{\label{fig:a2.08}
\includegraphics[scale=0.25,angle=270]{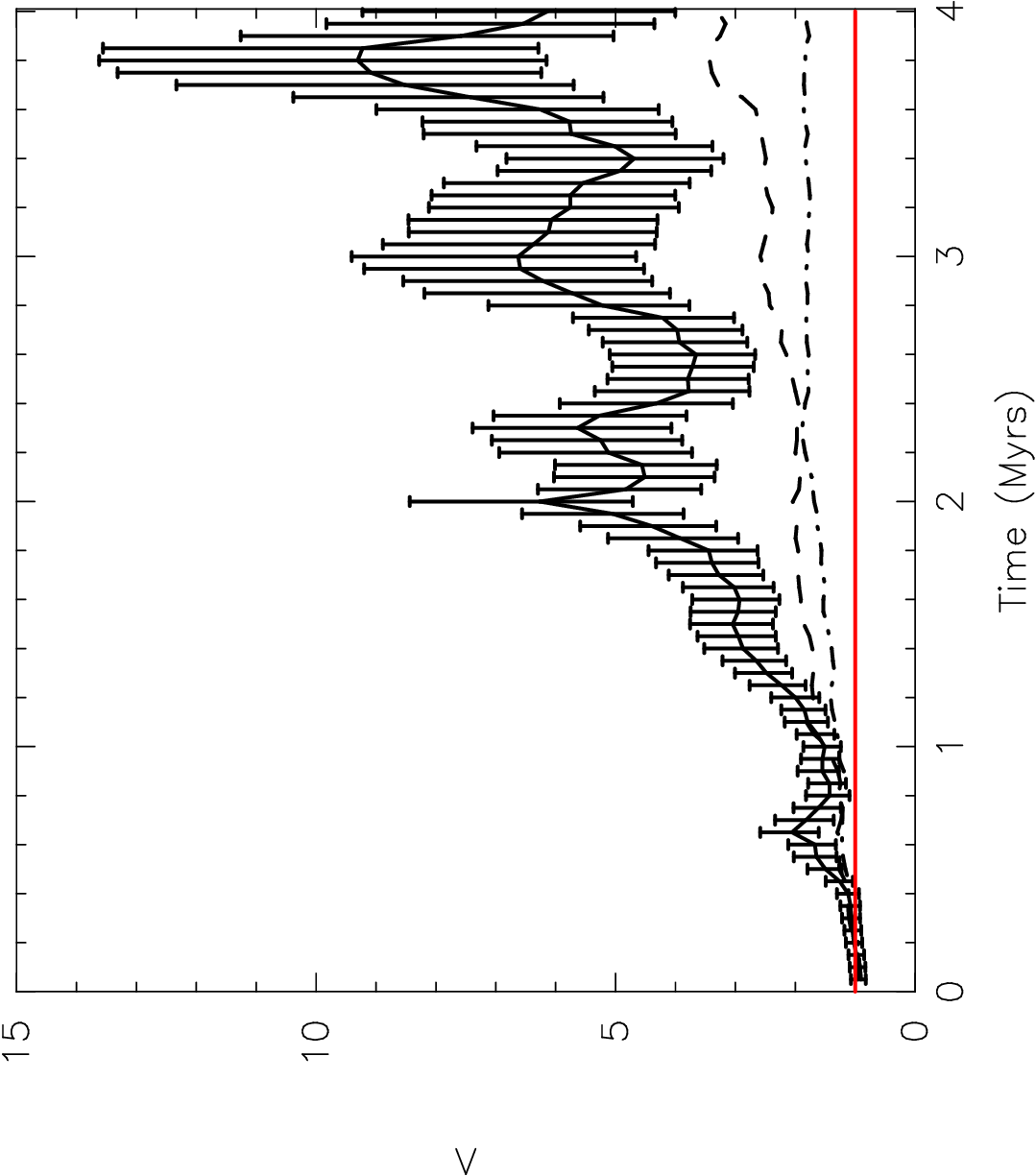}}
\subfigure[$D=2.0; Q=0.4$ -- b2.08]{\label{fig:b2.08}
\includegraphics[scale=0.25,angle=270]{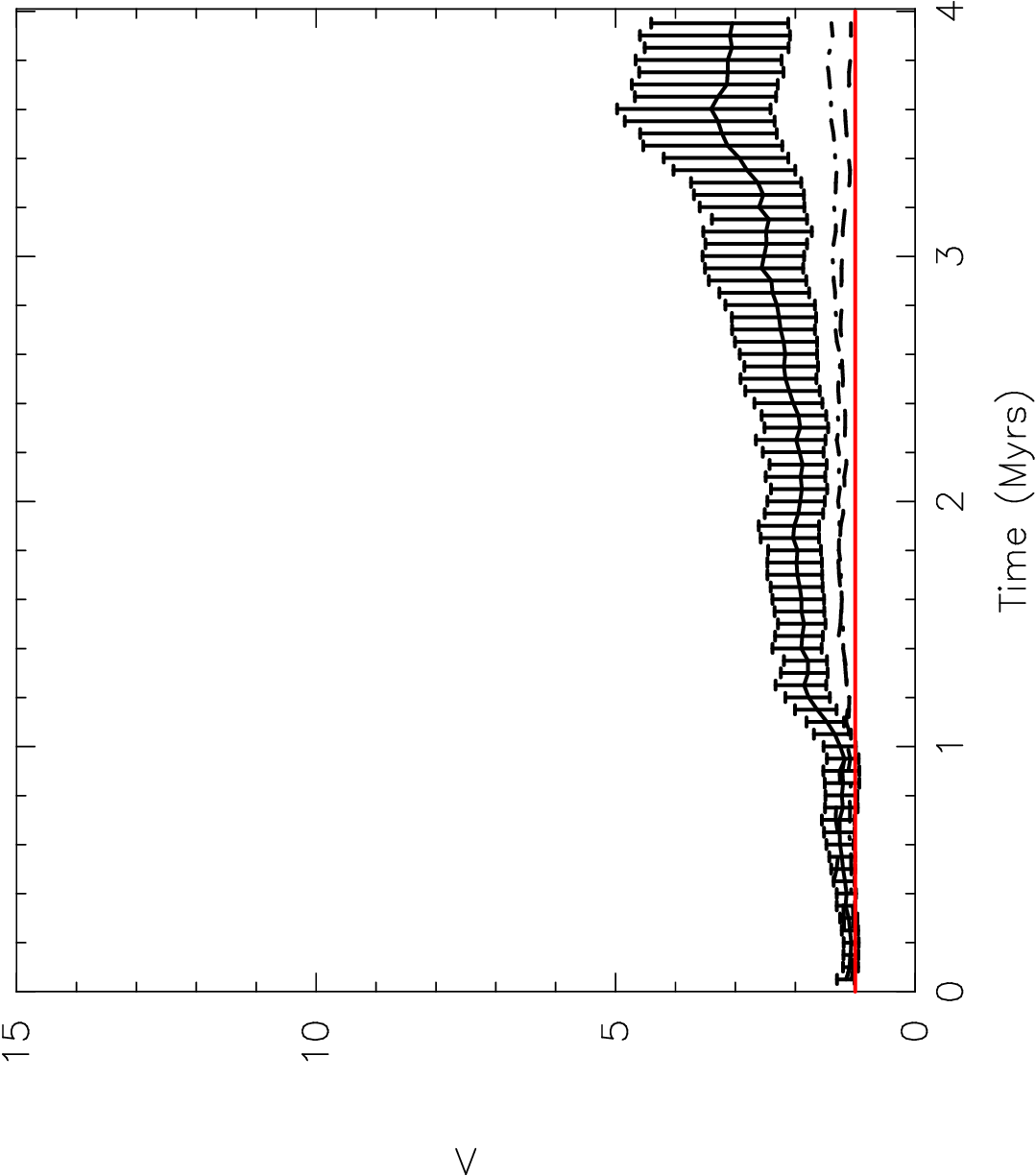}}
\subfigure[$D=2.0; Q=0.5$ -- c2.06]{\label{fig:c2.06}
\includegraphics[scale=0.25,angle=270]{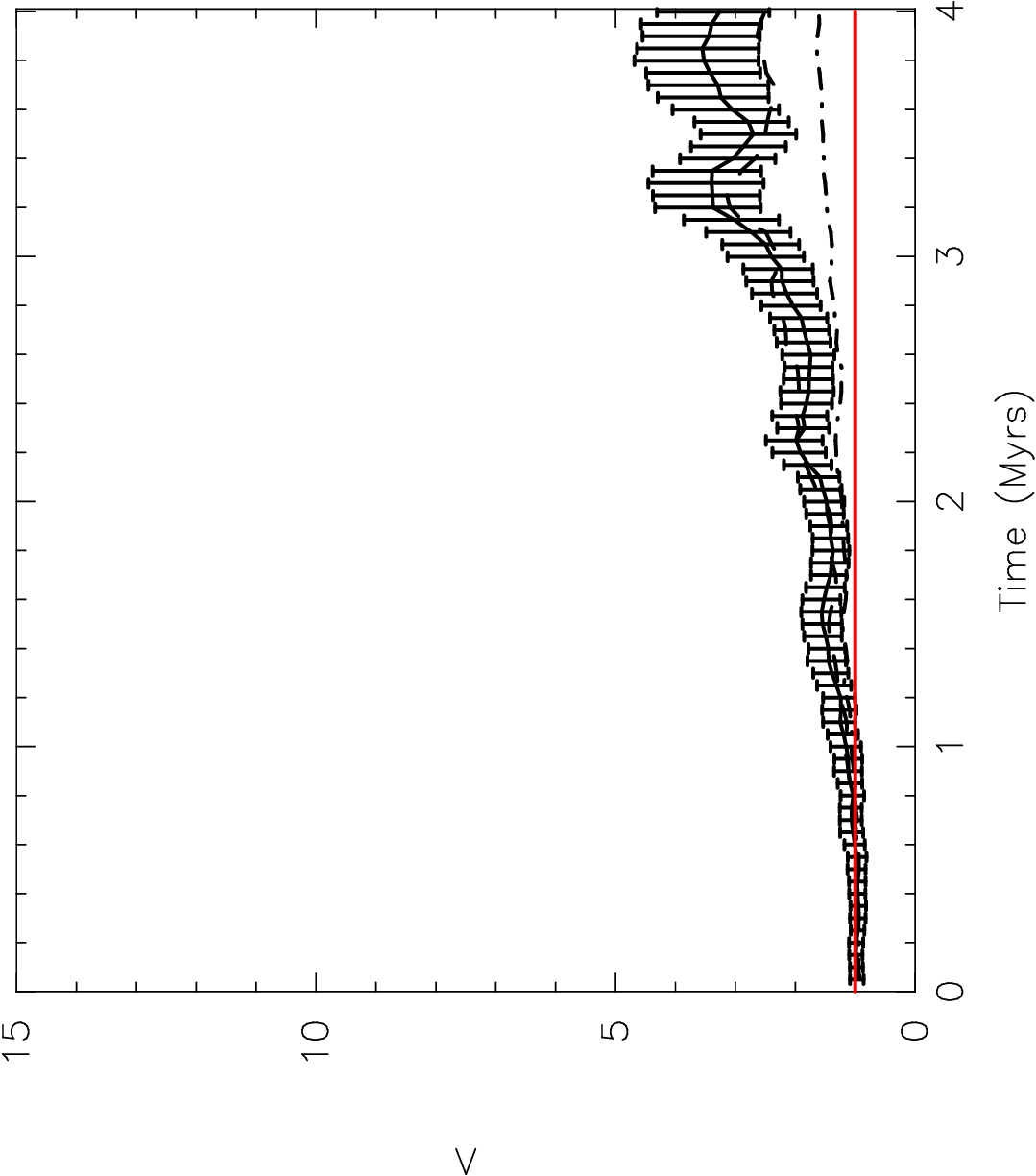}}

\subfigure[$D=2.6; Q=0.3$ -- a3.04]{\label{fig:a3.04}
\includegraphics[scale=0.25,angle=270]{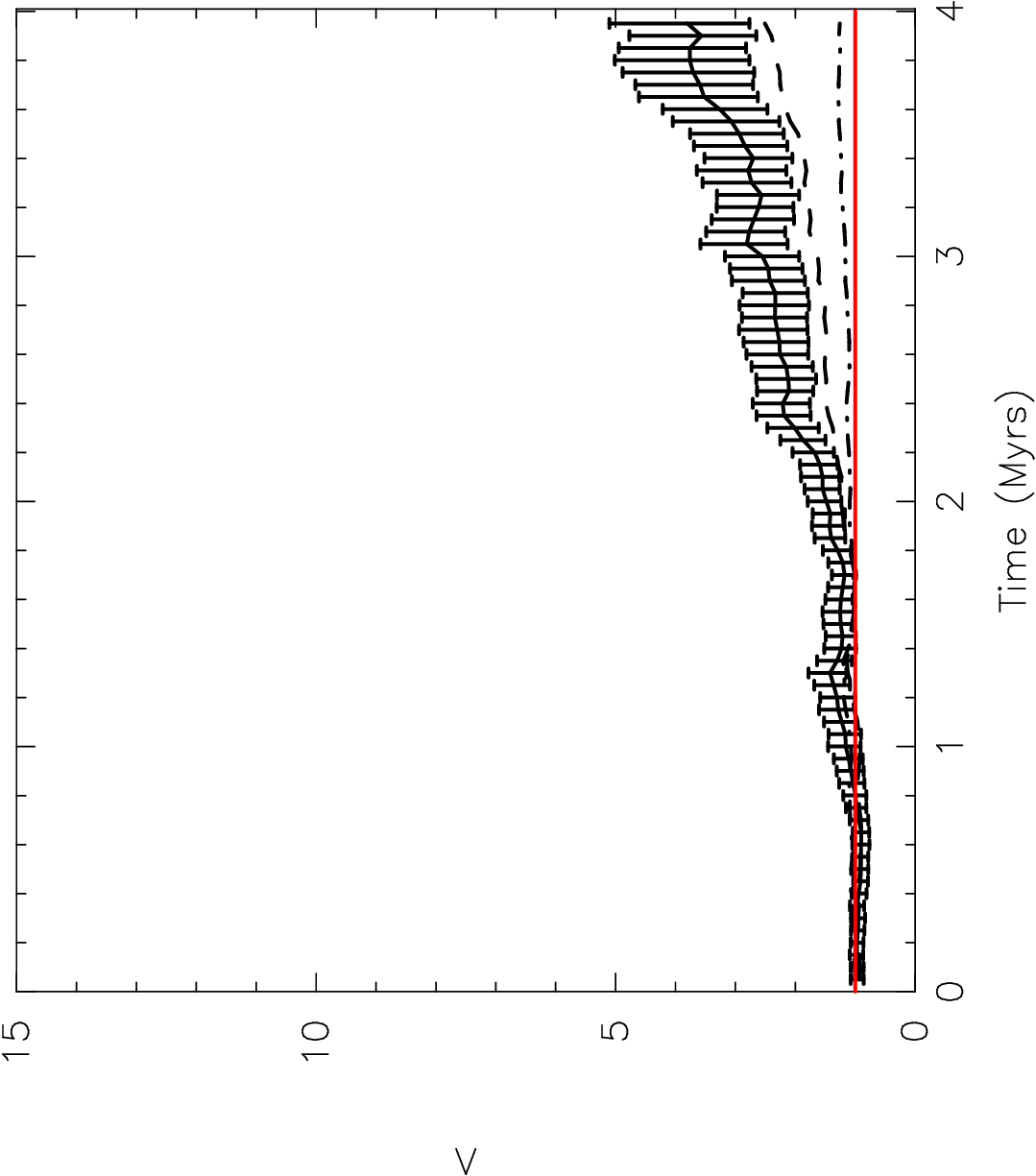}}
\subfigure[$D=2.6; Q=0.4$ -- b3.02]{\label{fig:b3.02}
\includegraphics[scale=0.25,angle=270]{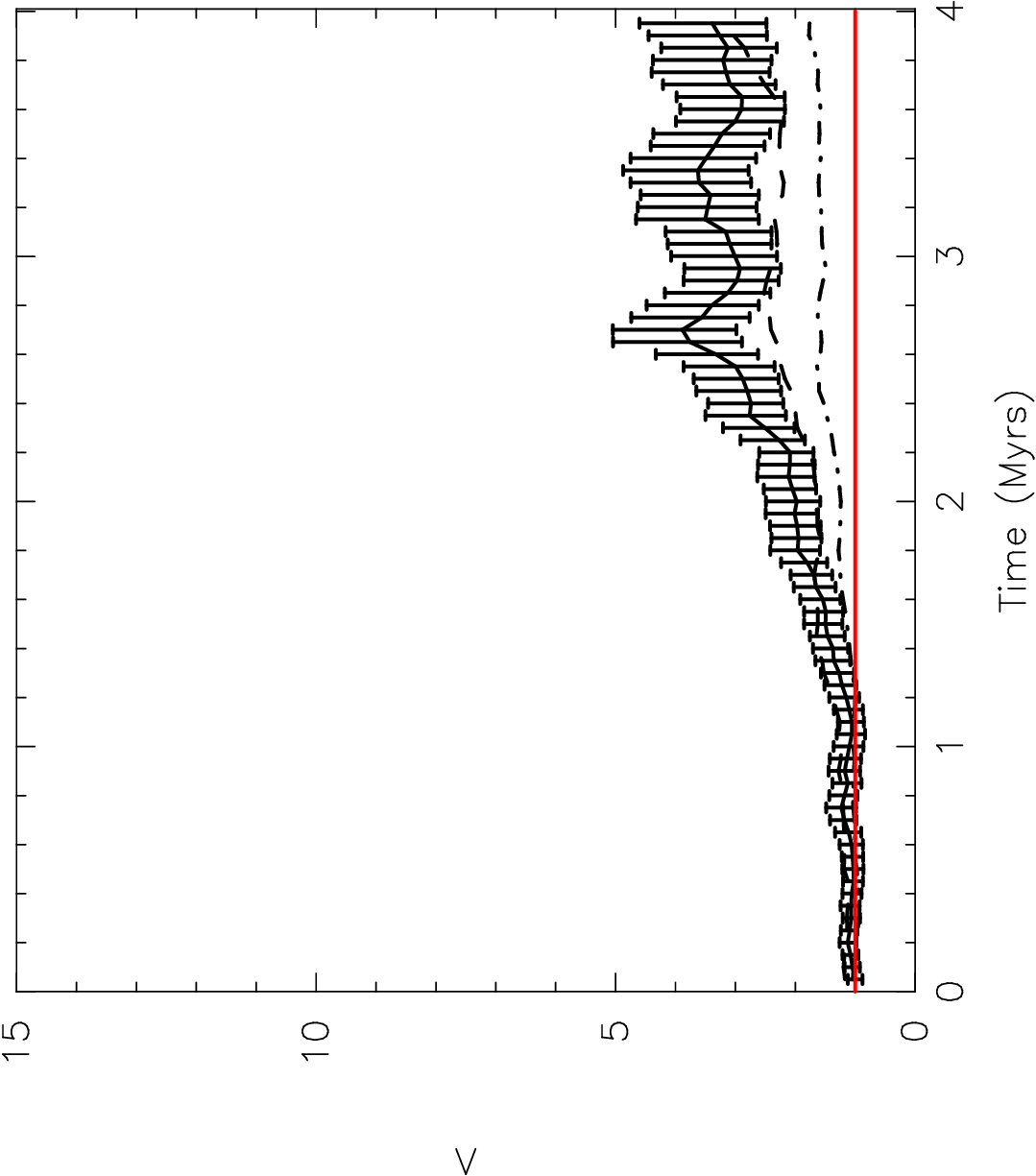}}
\subfigure[$D=2.6; Q=0.5$ -- c3.09]{\label{fig:c3.09}
\includegraphics[scale=0.25,angle=270]{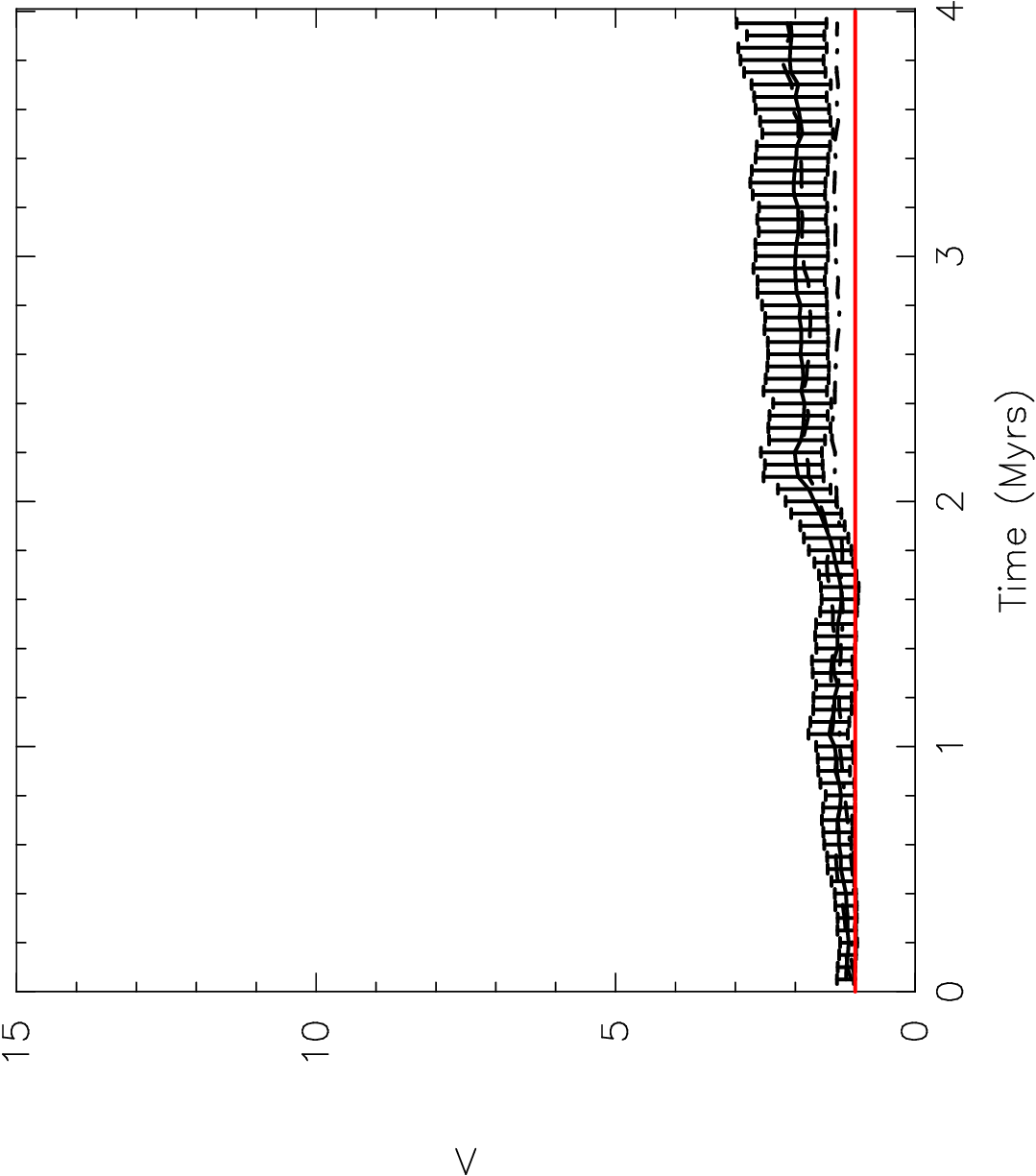}}

\subfigure[$D=3.0; Q=0.3$ -- a4.02]{\label{fig:a4.02}
\includegraphics[scale=0.25,angle=270]{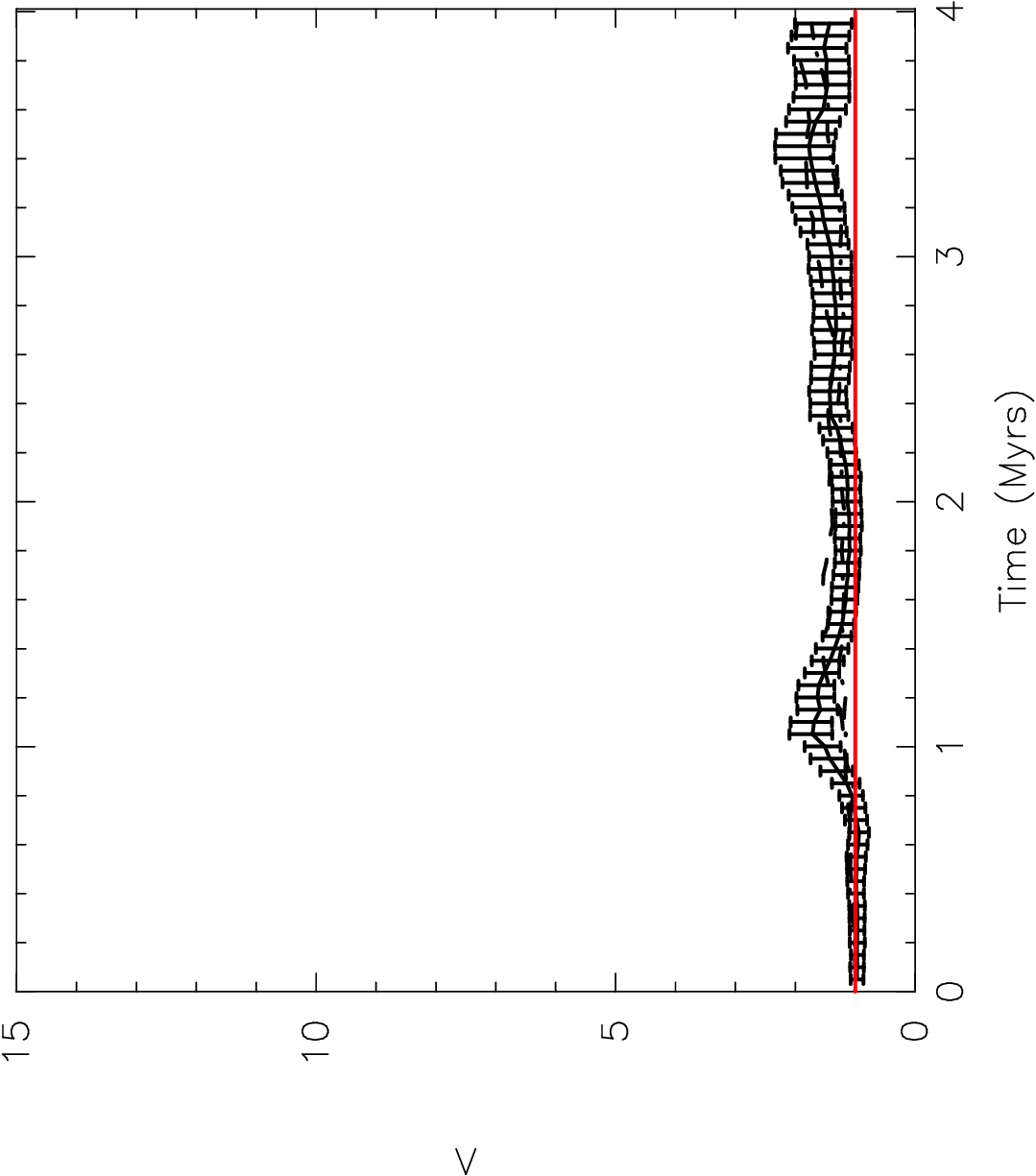}}
\subfigure[$D=3.0; Q=0.4$ -- b4.03]{\label{fig:b4.03}
\includegraphics[scale=0.25,angle=270]{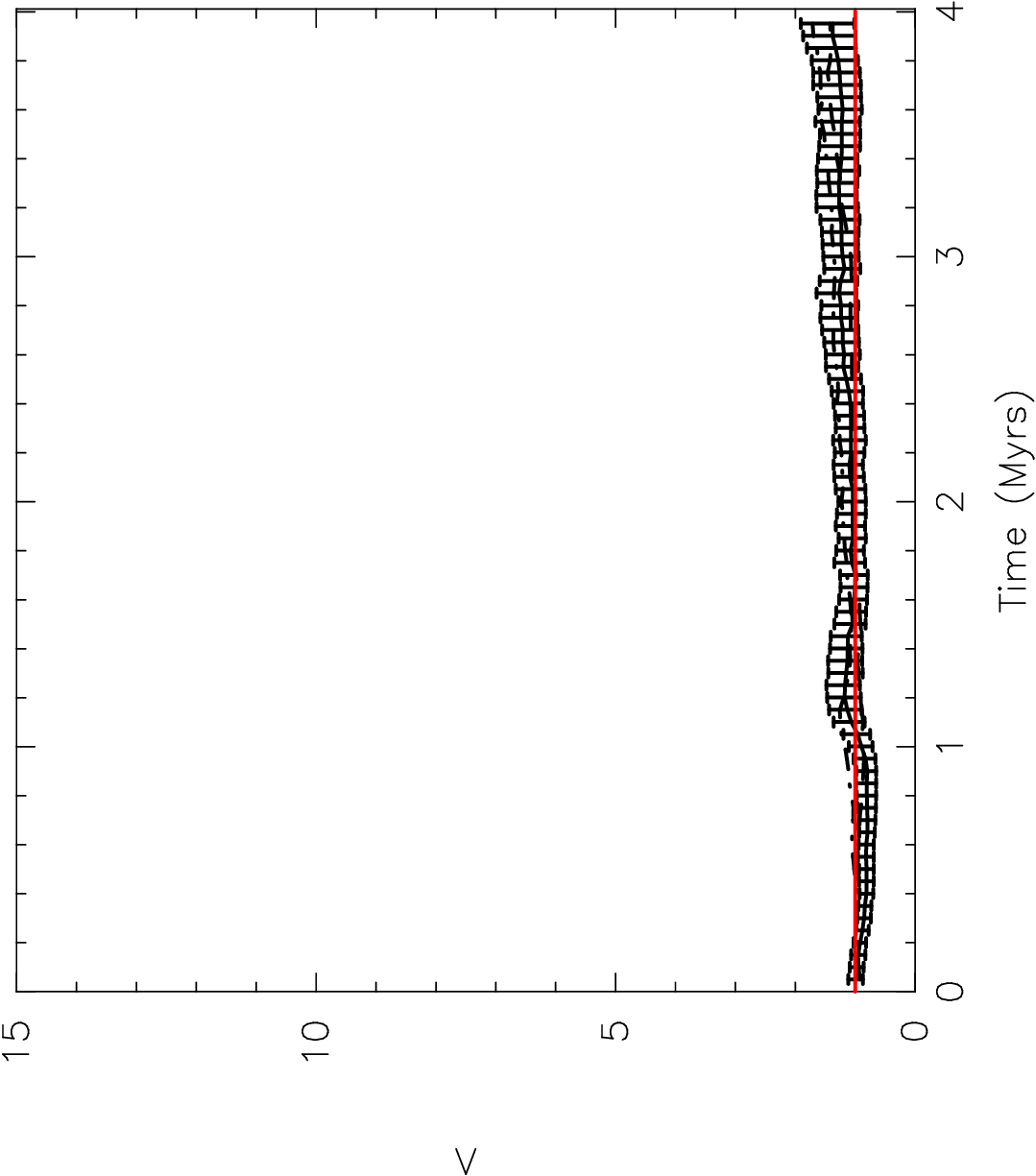}}
\subfigure[$D=3.0; Q=0.5$ -- c4.03]{\label{fig:c4.03}
\includegraphics[scale=0.25,angle=270]{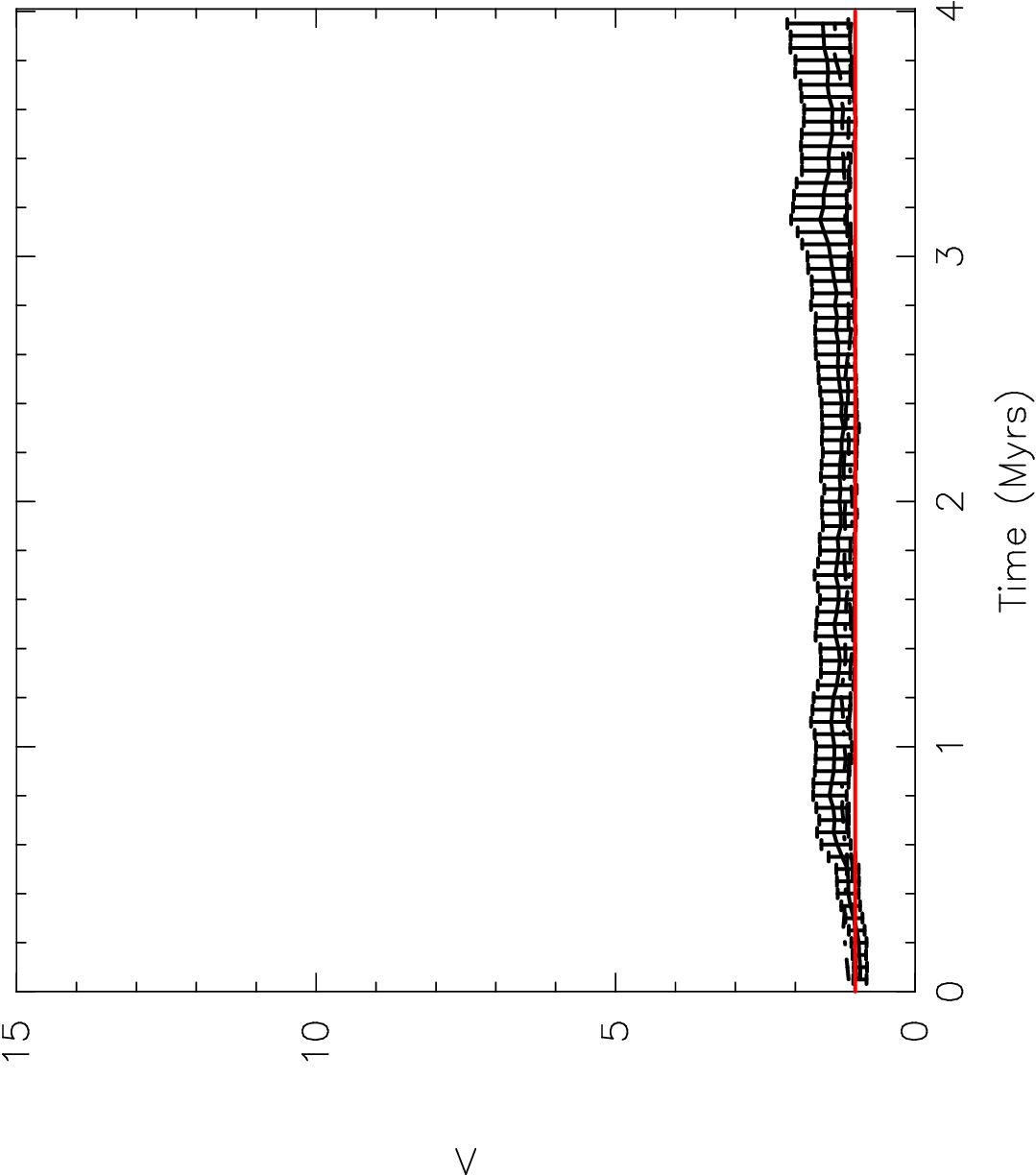}}

%\hspace*{0.6cm}
  \end{center}
  \caption[bf]{Generic examples from the investigated $D$ and $Q$
    values. Plots show the evolution of mass segregation ($\Lambda$)
    with time for $N=10$ (\textit{solid}, includes error bars); $N=20$
    (\textit{dashed}); $N=50$ (\textit{dot-dash}). The red line
    indicates a $\Lambda$ of unity i.e. no mass segregation. The error
    bars show a 1$\sigma$ deviation. A movie showing the evolution of
    (a) a1.21 can be downloaded at
    http://www.astro.group.shef.ac.uk/stars.html}
  \label{fig:DandQ}
\end{figure*}

In Fig.~\ref{fig:DandQ} we show the typical evolution of $\Lambda$
with time for clusters with different initial virial ratios and
fractal dimensions for $N=10,20 {\rm ~and~} 50$, the full
version of this figure, showing the evolution of all of the
simulations, can be found in the supplementary data.  In the top left
is our canonical $Q=0.3$, $D=1.6$ cluster from paper I.  From left to
right, the initial virial ratio increases from $Q=0.3$ to 0.4 and 0.5
(virialised).  From top to bottom, the fractal dimension increases
from $D=1.6$ (very clumpy) to $2.0, 2.6$ and 3.0 (roughly a uniform
density sphere).

It would be expected from our earlier argument that lower-$Q$ and
lower-$D$ clusters will collapse to a denser state, and hence show
more rapid and more pronounced dynamical mass segregation.  This is
exactly what is seen in Fig.~\ref{fig:DandQ}: faster and more intense
mass segregation at the top left (cool and clumpy clusters), and no
appreciable mass segregation at all at the bottom right (virialised,
uniform density clusters). The trend in the typical evolution of
clusters in our parameter space shown in Fig.~\ref{fig:DandQ} is
exactly what we would expect to see following the theoretical argument
in Section~\ref{ssec:dyn_ms}.

Whilst the behaviour of `typical' clusters is exactly what is
expected, many individual clusters show unusual and unexpected
behaviour. Examination of each simulation (see the supplementary data)
shows that, whilst the general pattern of evolution with $D$ and $Q$
holds, there is a large degree of stochasticity due to each fractal
being different in detail to every other fractal.

%----fig:DandQ----------
%NEED TO IDL THESE IMAGES??
%----fig:DandQ----------

\subsection{Stochasticity}
\label{ssec:stocasticity}

The supplementary data shows that a number of simulations do not show
the behaviour that we might expect for their $D$ and $Q$ values.  The
evolution of fractal clusters depends significantly on the specific
initial conditions of the cluster. The systems are inherently chaotic
and so each new random number seed can deliver completely different
evolution. Phenomena which occur during the evolution of a cluster
(such as mass segregation) can also be very transient. In some
systems, $\Lambda$ may change by a large factor on short
timescales. In this section we have chosen a few examples which
demonstrate the stochasticity and transience in the evolution of cool,
clumpy clusters.

\begin{figure}
  \begin{center}
    \setlength{\subfigcapskip}{10pt}
\subfigure[]{\label{fig:faa11a020.3}
\includegraphics[scale=0.3,angle=270]{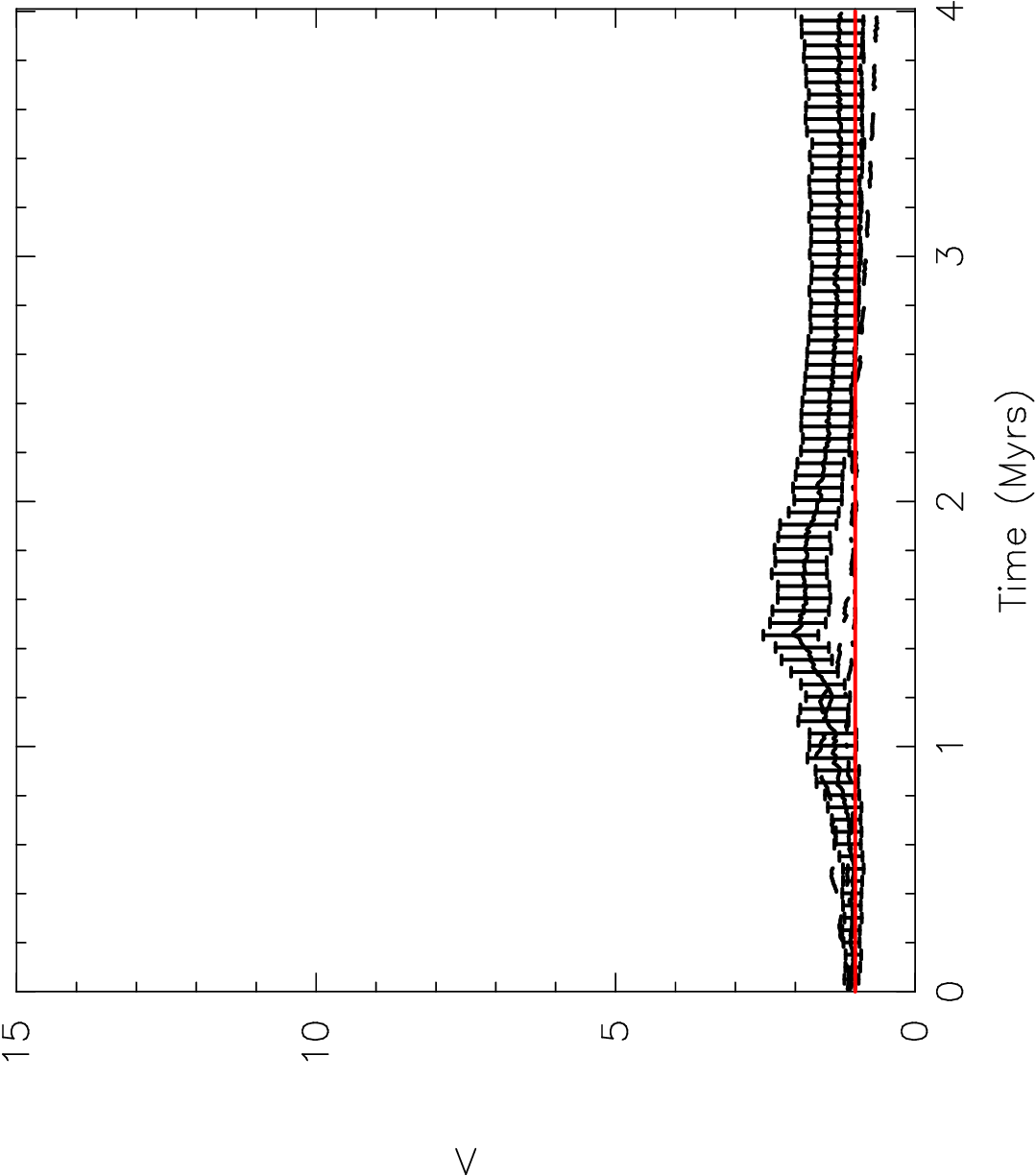}}
\subfigure[]{\label{fig:faa11a020.ext}
\includegraphics[scale=0.3,angle=270]{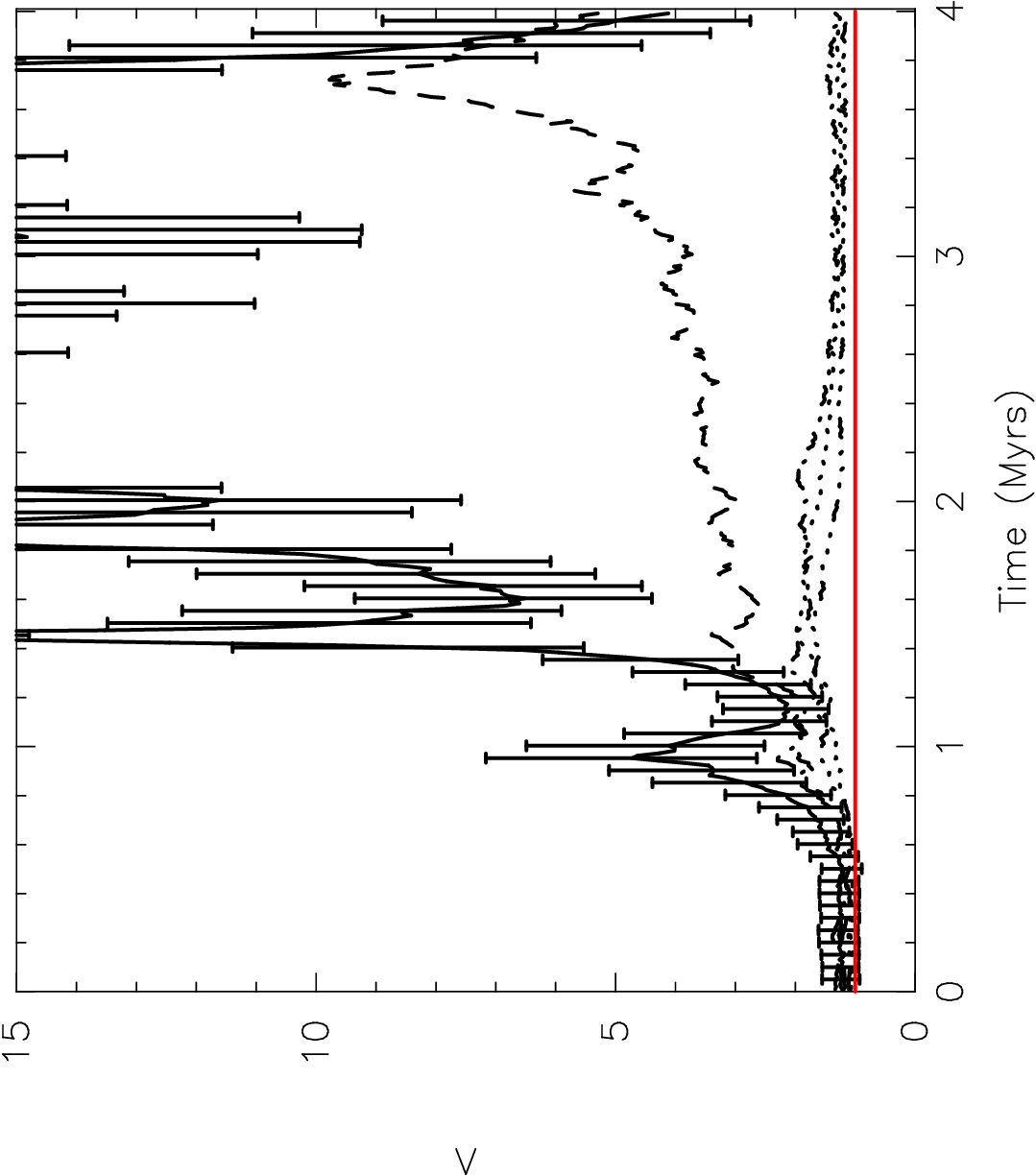}}
  \end{center}
  \caption[bf]{Evolution of $\Lambda$ with time for run
    a1.20. Top: $N=10,20 {\rm ~and~} 50$. Bottom: $N=4 {\rm~(solid~
      line)}, 6, 8, 10, {\rm ~and~}12$ (top to bottom) for run
    a1.20. Error bars have only been included for $N=4$ for clarity.
  \label{fig:faa11a020}}
\end{figure}

There are several $D=1.6, Q=0.3$ clusters which appear to have little
or no mass segregation, for example run a1.20 (see
Fig.~\ref{fig:faa11a020}). In this cluster there appears to be no
significant mass segregation at any time during the 4 Myr of the
simulation for $N=10,20 {\rm ~and~}
50$. Fig.~\ref{fig:faa11a020.ext} shows the evolution of $\Lambda$ for
run a1.20 in more detail than Fig.~\ref{fig:faa11a020.3}. In this plot
we show $\Lambda$ for $N=4, 6, 8, 10 {\rm ~and~} 12$. It is
now clear that the cluster {\em does} mass segregate, but only for
stars more massive than the 6\nth~most massive star. In fact, a
substantial multiple system consisting of the 4 most massive stars is
formed in this cluster. Mass segregation is only present for the
6\nth~most massive stars because the 7\nth~most massive star is
ejected early in the simulation, thereby enlarging the length of all
MSTs that include more than 7 stars. Mass segregation for $N=4$ occurs
at 0.8 Myr with $\Lambda = 3.4^{+1.8}_{-1.5}$, and by 2 Myr
$\Lambda$ for the four most massive stars has risen to
$50.9^{+22.0}_{-19.9}$, with a total MST length of 13000 AU (an
average separation of 3250 AU), $\Lambda$ is very high here because
many lower mass stars have been widely spread from the cluster, making
the random MSTs large. This also explains the large error values.
Fig.~\ref{fig:faa11a020.ext} illustrates the rapid variations in
$\Lambda$ for the small-$N$ system of the most massive stars, and the
fairly smooth variation for larger $N$.

Superficially, run c4.10 looks very similar to run a1.20, for $N=10,20
{\rm ~and~} 50$, in that it shows no evidence for mass segregation
throughout the simulation. When we investigate this cluster in detail,
we find that the two simulations are completely different as this
cluster does not undergo a dense collapse phase. It becomes only
slightly mass segregated at 3.2 Myr with $\Lambda$ for $N=4$ of $
3.1^{+1.5}_{-1.1}$.

%----fig:faa11a020----------
%----fig:faa11a020----------

\begin{figure}
  \begin{center}
    \setlength{\subfigcapskip}{10pt}
\subfigure[]{\label{fig:faa11a008.190}
\includegraphics[scale=0.37,angle=270]{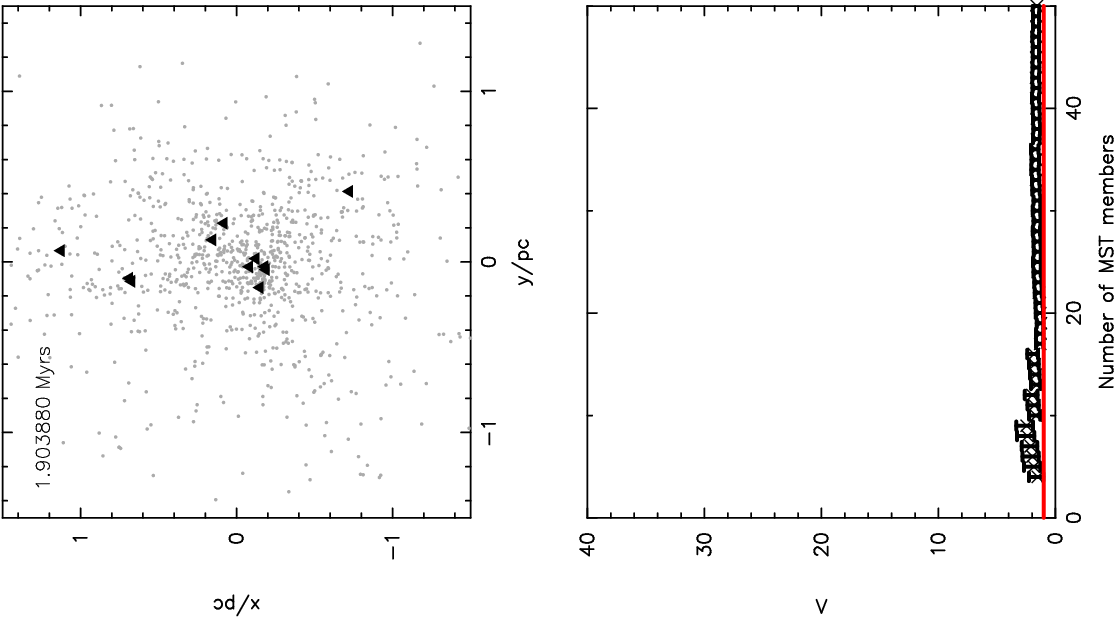}}
\subfigure[]{\label{fig:faa11a021.191}
\includegraphics[scale=0.37,angle=270]{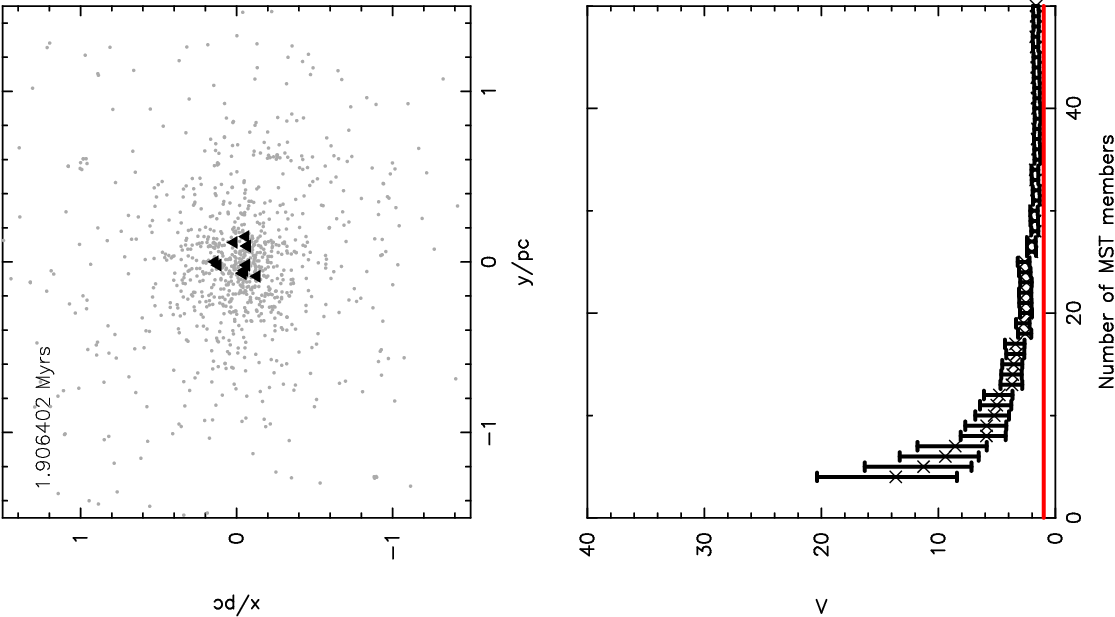}}
  \end{center}
  \caption[bf]{Runs a1.08 (a) and a1.21 (b): $D=1.6,Q=0.3$. The ten
    most massive stars are depicted by triangles. The most massive
    stars are not always clearly seen because of close grouping.
  \label{fig:faa11a008}}
\end{figure}

In runs a1.08 and a1.21 the stochasticity of runs with different
random number seeds is easily seen. Both simulations are initially
{\em statistically} the same, but in Fig.~\ref{fig:faa11a008} we show
the states of the clusters at $\approx 1.9$ Myr.  Run a1.08 shows no
evidence of mass segregation, whilst run a1.21 shows very significant
mass segregation out to $N = 20 - 25$ (in
Fig.~\ref{fig:faa11a021.191} the 10 most massive stars are very
closely clustered in the centre, and the symbols indicating their
positions overlap somewhat).

%----fig:faa11a008/21----------
%----fig:faa11a008/21----------

It is important to note that the plots used in Fig.~\ref{fig:DandQ} do
not show all of the mass segregation information. They only show mass
segregation for the $N$ which is plotted (in this case $N=10,20{\rm
  ~and~} 50$).  This choice is rather arbitrary and fails to provide
all of the information.  We have chosen to plot these values for
clarity, but any detailed investigation of the evolution of cool,
clumpy clusters requires an analysis of a three dimensional
dataset including time, $N$, and $\Lambda$.

\begin{figure}
  \begin{center}
    \setlength{\subfigcapskip}{10pt}
\subfigure[]{\label{fig:faa11a004.212}
\includegraphics[scale=0.37,angle=270]{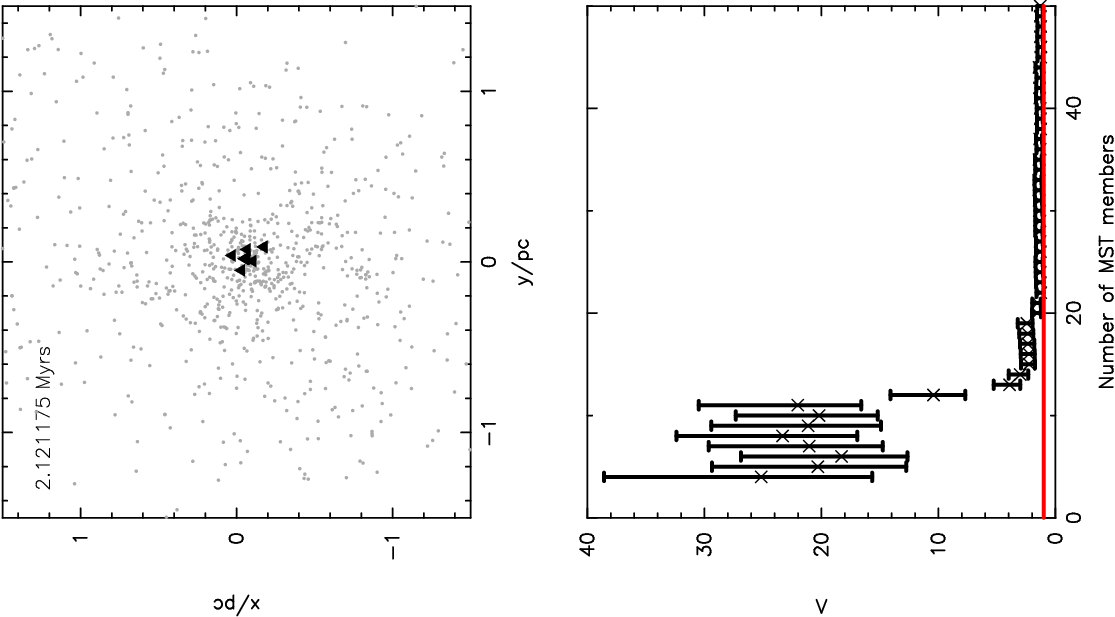}}
\subfigure[]{\label{fig:faa11a004.222}
\includegraphics[scale=0.37,angle=270]{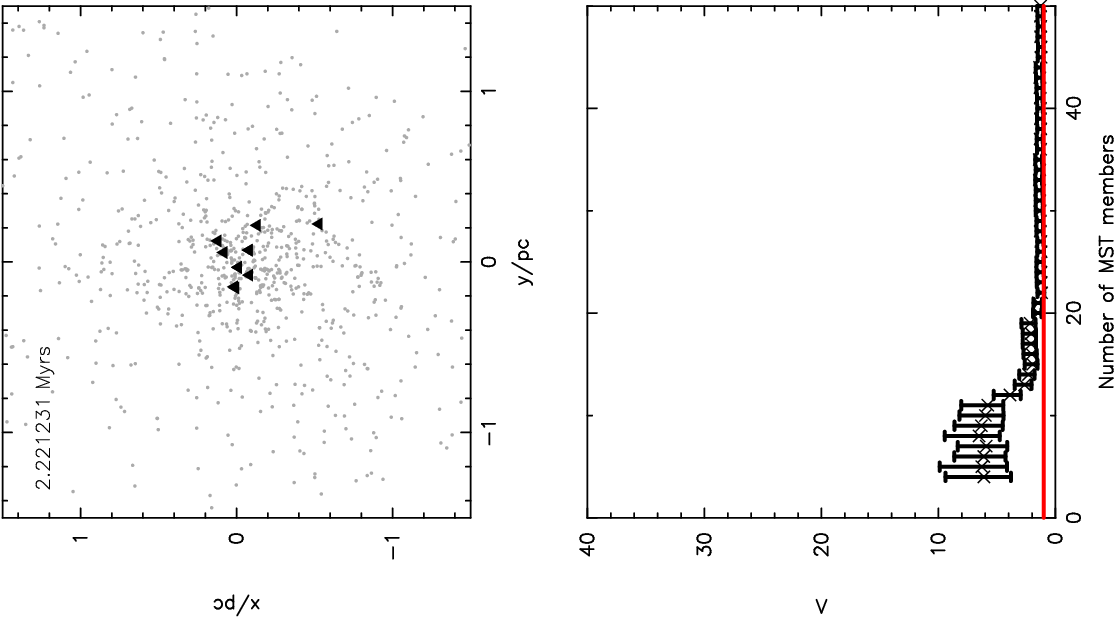}}
  \end{center}
  \caption[bf]{Run a1.04: $D=1.6,Q=0.3$. The ten most massive stars
    are depicted by triangles. (a) At 2.1 Myr the cluster shows very
    high levels of mass segregation for the twelve most massive stars
    ($\Lambda\approx 21$). The ten most massive stars are tightly
    grouped in the centre of the the cluster, and cannot be seen
    clearly. (b) 0.1 Myr later, the amount of mass segregation present
    in the cluster has been vastly reduced ($\Lambda\approx 6$). There
    are two massive star-massive star binaries in the central region,
    causing symbols to overlap.
  \label{fig:faa11a004}}
\end{figure}

In run a1.04, the stochastic and transient nature of mass segregation
can be seen. Fig.~\ref{fig:faa11a004.212} shows that at 2.1 Myr the
twelve most massive stars in the cluster are in a heavily mass
segregated state. However, Fig.~\ref{fig:faa11a004.222}, shows that
only 0.1 Myr later the amount of observed mass segregation in the
cluster has been reduced from $\Lambda \approx 21$ to $\Lambda \approx
6$ because of the disruption of the massive multiple system at the core.

%----fig:faa11a004----------
%----fig:faa11a004----------

As observations are a snapshot of one time in a cluster's evolution,
they only provide two dimensions of $N$ and $\Lambda$.  This means
that observations miss the time evolution of the system which, as is
clear in many plots in the supplementary material, can vary extremely
rapidly.  Hence we stress that observations only provide a snapshot of
a young cluster. {\em Young clusters can evolve very rapidly} on a
timescale of $\sim$ 1 Myr. Therefore the instantaneous state of a
cluster (especially if it is based on two dimensional positions) may
provide no information on what the cluster was like in the recent
past, or to how it might evolve in the near future \citep[see
  also][]{goodwin06,bastian08,allison09b}.

%----fig:faa11a002----------
\begin{figure}
  \begin{center}
    \setlength{\subfigcapskip}{10pt}
\subfigure[]{\label{fig:faa11a002.185}
\includegraphics[scale=0.3,angle=270]{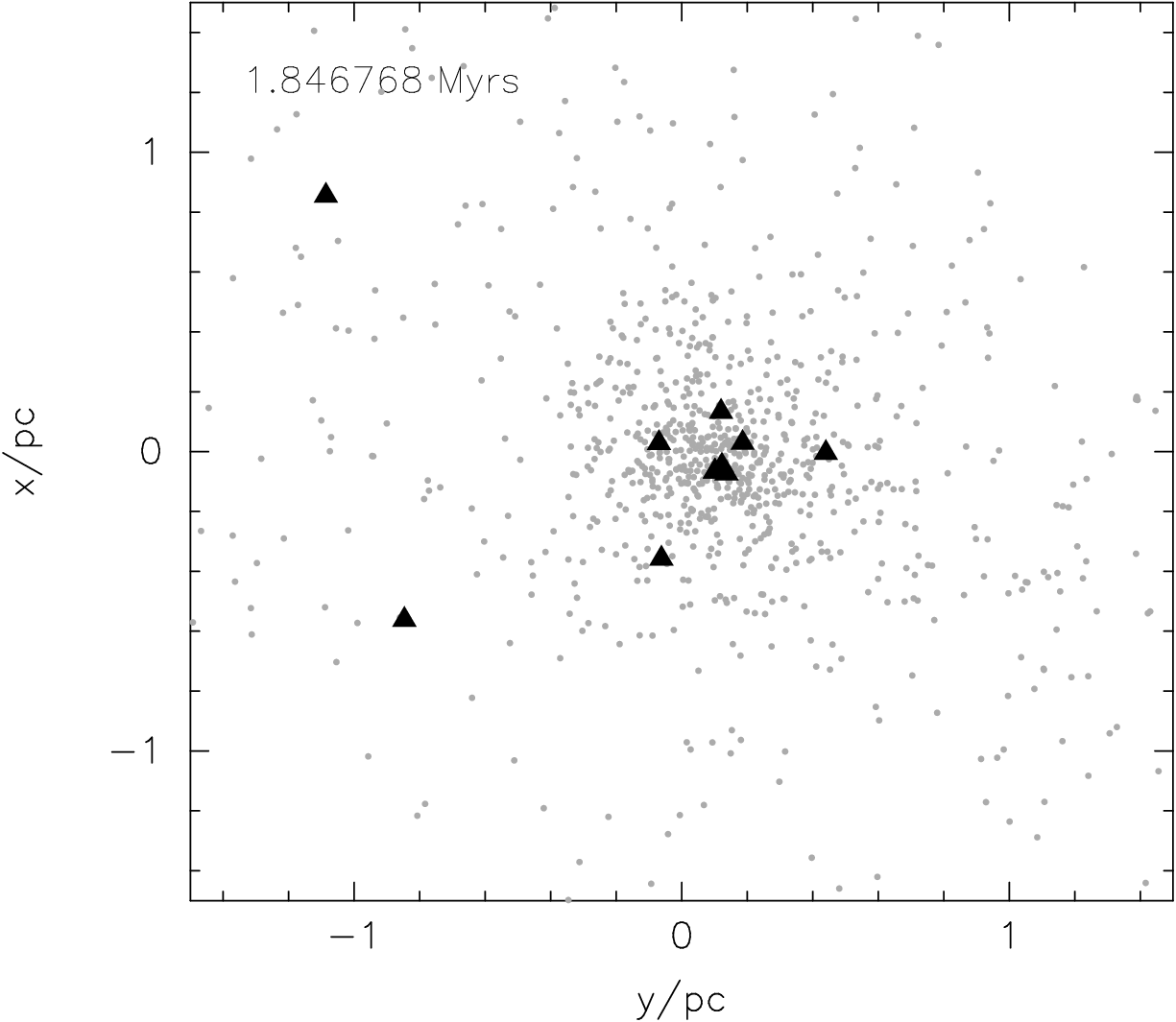}}
\subfigure[]{\label{fig:faa11a002.250}
\includegraphics[scale=0.3,angle=270]{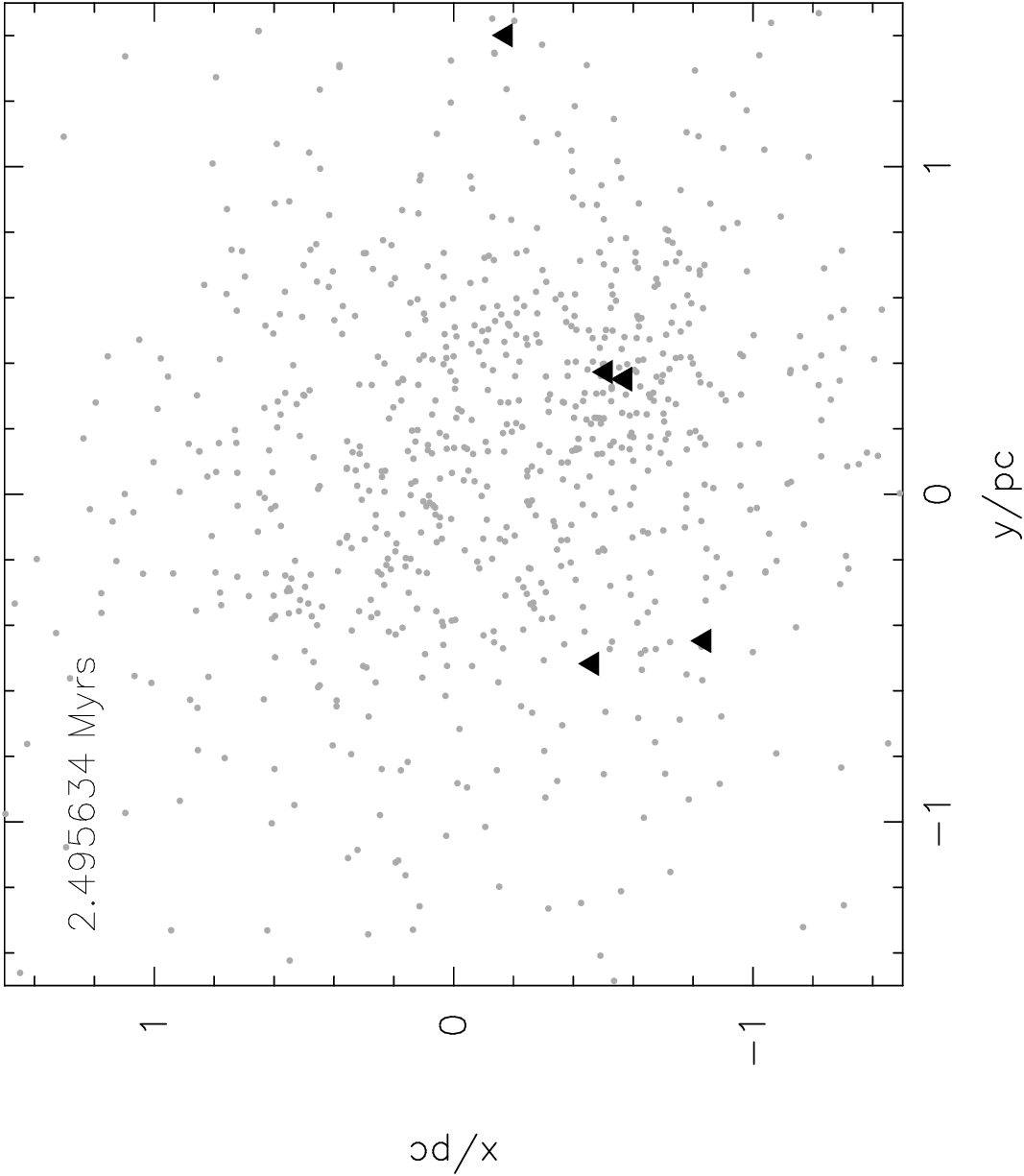}}
  \end{center}
  \caption[bf]{Run a1.02: $D=1.6,Q=0.3$. The ten most massive stars
    are depicted by triangles. (a) The cluster has evolved to form a
    centrally concentrated distribution. In the core there is a tight
    triple system consisting of 3 of the most massive stars, which
    cannot clearly be seen. (b) 0.7 Myr after (a), the cluster has
    been destroyed by the decay of the multiple system, and most of
    the 10 most massive stars have been ejected.
  \label{fig:faa11a002}}
\end{figure}
%----fig:faa11a002----------

%-----------------------------------------------------------------------
\subsection{Post-collapse evolution}
\label{ssec:post-collapse}

The collapse of clumpy clusters may not lead only to mass segregation,
but also to other interesting phenomena; most notably the formation of
`Trapezium-like' systems, the rapid ejection of massive stars, and the
disruption of the cluster itself.

\subsubsection{Formation of `Trapezium-like' systems}
\label{sssec:trapezium}

The Trapezium in the Orion Nebula Cluster (ONC) is a multiple system
consisting of four of the most massive stars in the ONC. For the
following discussion we shall use a simple definition of a
`Trapezium-like' system as a close group of massive stars (at least 3
with $M>8\Msun$) which are bound.  Under this criterion, 45 of the 50
$D=1.6,Q=0.3$ simulations formed a `Trapezium-like' system. These
simulations contained no primordial binaries, but were able to migrate
the most massive stars into very close proximity to each other on very
short timescales ($\sim$ 1 Myr) through dynamics alone \citep[see
  also,][]{allison09b}. In the 50 $D=1.6,Q=0.3$ simulations, the
average shortest separation reached by the four most massive stars was
0.04~pc (about 8000~AU), which occurs within the first 2 Myr. The
shortest separation between the four most massive stars in any
simulation was 0.0025~pc (500~AU). We would expect that with different
initial conditions (i.e., initial binaries, smaller $R_0$, etc) these
multiple systems could be made even more extreme. In fact, preliminary
investigations of simulations with an initial fractal radius of 0.5~pc
(half the size used in this paper) and only single stars found a
`Trapezium-like' system with the four most massive stars in a cluster
with an average separation of only 20~AU between them.

`Trapezium-like' systems form during the dense core phase as the most
massive stars loose kinetic energy through two-body encounters with
lower-mass stars.  As they mass segregate, the typical separations
between the most massive stars decrease and they form high-order
multiple systems.  We note that the dense core that forms at the peak
of the collapse is similar to the initial conditions used by
\citet{pflamm-altenburg06} to model the formation of the Trapezium in
the ONC.  It also has the extreme density required to explain
observations of $\eta$ Cham \citep{moraux07} and the ONC
\citep{parker09}.  We will examine the formation and evolution of
`Trapezium-like' systems in more detail in a subsequent paper
(R.~J.~Allison et al., in prep.). Hydrodynamical simulations also
  show the formation of trapezium-like systems in the gas-dominated
  phase of cluster formation
  \citep[see,][]{bonnell03,klessen09}. These systems are able to form
  due to the ability of gas to dissipate kinetic energy and
  redistribute angular momentum. Therefore, if the effects of gas were
  included in these simulations we would expect the trapezium systems
  which form in these simulations to appear earlier and to be even
  more extreme.

\subsubsection{Core collapse?}

Some simulations appear to show core collapse events.  Early mass
segregation can create a core-halo structure in the cluster, with a
dense core of massive stars, surrounded by a halo of lower-mass stars.
The dense core often forms a Trapezium-like system (see above).  The
lower-mass stars are typically on radial orbits, and may enter and
extract energy from the core causing it to shrink and the higher-order
massive multiples to harden.  This also causes the halo to heat up and
the cluster to start dissolving.  At some point, the massive multiples
may decay, ejecting high-mass stars at high velocity (see below) and
blowing the cluster apart.

In run a1.02, for example, the cluster collapses and forms a dense
core. The 4 most massive stars form a Trapezium-like system which very
rapidly decays and destroys the cluster. Fig.~\ref{fig:faa11a002.185}
shows the cluster during the `core phase', the massive star multiple
system is in the centre of the cluster (but the individual stars are
too close to resolve in this image).  Fig.~\ref{fig:faa11a002.250}
shows the cluster 0.7 Myr later. Most of the massive stars have been
ejected from the central 1 pc, and the cluster has been destroyed.
Prior to dissolution, the degree of mass segregation increases as the
central multiple hardens, and at around 1.9~Myr the central multiple
finally decays, ejecting the most massive stars and creating a hard
binary. This system contains two stars of mass 45.3 and 27.8\Msun,
which had an initial separation of 800 AU before the ejection, and a
separation of $\sim 100$ AU after. This binary increases its binding
energy by around $1\times 10^{40}$ J, which is comparable to the
entire potential energy of the cluster of $\sim -6\times 10^{39}$
J. In this simulation the formation of the binary system alone is
  able to directly disrupt the cluster. It is also possible that the
  energy input comes from gradual interactions with massive star multiples
  which slowly disrupt the cluster and form a hard massive star binary.

% Ok, the numbers are rough for the binary, but close enough.
% Assuming 500\Msun for the cluster, and a plummer radius of 0.1 pc 
% (from images)

%----tab:disruption----------
\begin{center}
\begin{table}
  \begin{tabular}{|c|c|c|c|c|}
\hline
        &  \multicolumn{4}{|c|}{$D$}      \\
$Q$     &  1.6    &  2.0   &  2.6  &  3.0 \\
\hline
0.3     &  32/50  &  2/10  &  0    &  0   \\
0.4     &  6/10   &  2/10  &  0    &  0   \\
0.5     &  7/10   &  0     &  0    &  0   \\
\hline

  \end{tabular}
\caption{Fraction of clusters which become globally unbound with 
varying initial virial ratio, $Q$, and fractal dimension, $D$.
\label{tab:disruption}}
\end{table}
\end{center}
%----tab:disruption----------

Table~\ref{tab:disruption} shows the number of simulations for which
the final (4 Myr) virial ratio is greater than unity (i.e., the cluster is
unbound). Energy is conserved in the simulations, but an unbound
cluster may be formed from an initially sub-virial cluster through the
redistribution of energy in the core collapse phase.

Even if the cluster manages to survive for the 4~Myr of our
simulations, shortly after this the most massive stars will become
supernovae and the loss of the most bound portions of the cluster
should destroy the cluster.  Thus clusters can be destroyed
dynamically without the need for gas expulsion
\citep{goodwin06,goodwin09}.

\subsubsection{Massive star ejections}
\label{sssec:ejection}

%----fig:vel----------
\begin{figure}
  \begin{center}
    \setlength{\subfigcapskip}{10pt}
    \includegraphics[scale=0.35,angle=270]{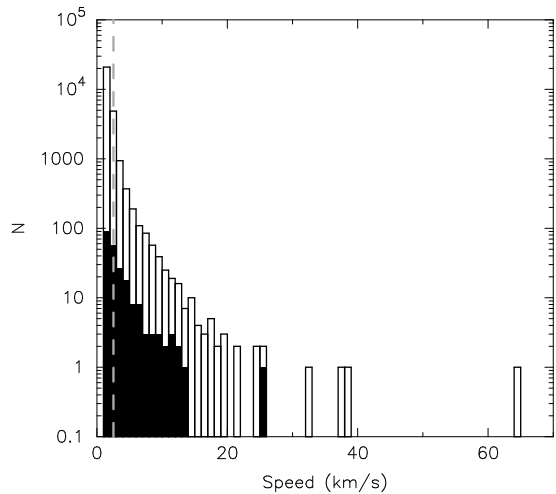}
  \end{center}
  \caption[bf]{Distribution of final speeds of all stars from
  the 50 $D=1.6,Q=0.3$ simulations. The black region shows stars with
  $M>10\Msun$, and the dashed line shows the average escape speed
  ($v_{\textrm{esc}}\approx 2.5$ km s$^{-1}$). 
  \label{fig:vel}}
\end{figure}
%----fig:vel----------

The ejection of massive stars from clusters can occur in three ways.
Firstly, Trapezium-like systems are inherently unstable and can
decay dynamically \citep{sterzik98}.  Secondly, other massive stars
can interact with Trapezium-like systems or with massive binaries and
be ejected.  Thirdly, the hardening of Trapezium-like systems in
core collapse processes may decrease its stability and cause it
to decay faster than it might otherwise have done.

Figure~\ref{fig:vel} shows the final (at 4 Myr) velocity distribution
of the 50 $D=1.6,Q=0.3$ simulations, for comparison the typical
velocity dispersion of a cluster is $\sim 1.9$~km~s$^{-1}$, and the
typical escape velocity is $\sim 2.5$~km~s$^{-1}$. The plot shows that
it is quite possible for the massive star ejections to reach
velocities in excess of 10 km s$^{-1}$. The simulations produce stars
with masses $>30\Msun$ having ejection velocities $>10$ km s$^{-1}$,
and in one case a 12\Msun star is ejected with a velocity of
25.9\kms. In all, 13 per cent of stars $>10$\Msun~ were ejected with
velocities $> 5$\kms ($\approx 5 {\rm~pc~Myr}^{-1}$).

\section{Discussion}
\label{sec:disc}

Many young clusters are observed to be `mass segregated', that is
their most massive stars are concentrated preferentially towards their
centres.  The ONC (2-3 Myr) is well known to be mass
segregated. \citet{hillenbrand98} find, using the cumulative
distribution of different mass groups, evidence of mass segregation
for stars more massive than 5\Msun, and a possibility that mass
segregation is present for stars $>$1-2\Msun. \citet{allison09} show,
using the minimum spanning tree method, that mass segregation in the
ONC is split into three `levels'. The first contains the four most
massive stars, the second stars $>5\Msun$, and the third is a decrease
towards no mass segregation below $5\Msun$.  Using the radial
variation of the IMF, \citet{harayama08} and \citet{sharma07} show
that NGC 3603 (2.5 Myr) and NGC 1893 (4 Myr) are mass segregated, but
that there is no evidence for mass segregation for stars $<1-2\Msun$
in either cluster.  \citet{raboud98} show, using the cumulative
distribution method, that NGC 6231 (4 Myr) is mass segregated. This
cluster, like the ONC, has different `levels' of mass segregation. The
most massive stars ($>16\Msun$) are clearly more mass segregated than
stars of lower mass, and stars with masses between $2.5 \textrm{ and }
16\Msun$ are spatially well mixed, but more mass segregated than stars
$<2.5\Msun$.  Also using the cumulative distribution method
\citet{jose08} find evidence for mass segregation in Stock 8 (1-5
Myr), and show that it is only apparent for stars $>1\Msun$. 
  Using our method, \citet{sana10} find that the cluster Trumpler 14
  ($\sim$4 Myr) also shows mass segregation. There is significant mass
  segregation for stars more massive than $\approx$10\Msun, but no
  evidence for stars less massive than this.

Such observations are a strong indication that these clusters have
undergone an early dense phase. The mass segregation in these clusters
follows a similar profile -- mass segregation is present for the most
massive stars, and below some particular mass ceases to be observed --
this is what we observe in our simulations.

There are two important elements that are missing from our
simulations: binaries and gas.  Binaries have been neglected as we
wish to explore the basic gravitational physics which binaries would
confuse.  In future papers, we will introduce primordial binaries into
our simulations.  Gas has been ignored, as it is computationally
extremely expensive to include and because introducing gas introduces
a whole new set of parameters (the equation of state of the gas, the
Mach number and power spectrum of the turbulence, etc.).  But clearly
gas is a vital ingredient in very young star clusters. It is what the
stars form from, and it can be a significant contributor to the
background potential.

From our simulations, it is not clear if massive stars might form
directly in massive cores \citep[e.g.,][]{krumholz07}, through
fragment-induced starvation \citep{peters10} or due to competitive
accretion \citep[e.g.,][]{smith09}.  However, hydrodynamical
simulations indicate that gravoturbulent fragmentation leads to
subclustered and mass segregated clusters
\citep{maclow04,maschberger10}. Clearly, the ability of young clusters
to dynamically mass segregate on a very short timescale means that
massive stars do not {\em have} to form (or rather, build up their
mass) in the core of a young cluster.  Massive stars can form in
subclusters or in the outskirts of clusters and within a~Myr or so be
very mass segregated. However, not only stars, but also gas, will be
involved in the collapse of the cluster to a dense phase and so cool
collapse might enhance competitive accretion.  Indeed, it is difficult
to imagine that some competitive accretion will not occur during the
channelling of stars and gas into a very dense state.  Do massive
stars form in massive cores that are then dynamically mass segregated?
Does competitive accretion dominate, especially during the dense
collapse to form the massive stars?  Do fairly massive stars form in
the outskirts of clusters which are dynamically mass segregated and
then have their masses increased by competitive accretion in a hybrid
scenario (such a process may be occurring in simulations; see
Maschberger et al. 2010)?

\section{Conclusions}
\label{sec:conc}

Observations and theory both strongly suggest that the initial
conditions of star clusters are cool and clumpy.  Therefore, we have 
conducted a large number of $N$-body simulations of the early
($< 4$~Myr) evolution of clusters with virial ratios of $Q = 0.3, 0.4$
and $0.5$ (where $0.5$ is virialised), and fractal dimensions of $D =
1.6, 2.0, 2.6$ and $3.0$ (where $3.0$ is roughly a uniform density
sphere) with radii of $1$~pc and $1000$ members (ie. total masses of $\sim
500$~M$_\odot$).   In our simulations, all  members were
initially single stars selected from a Kroupa IMF, and the simulations 
lasted for 4~Myr (so not requiring stellar evolution to be included).  

We study the evolution of the star clusters with a particular emphasis
on the level of mass segregation.  We measure mass segregation using a
minimum spanning tree to provide a quantitative measure of mass
segregation and which is not biased by clumpy underlying mass
distributions (Allison et al. 2009a).

This study follows that of Allison et al. (2009b), in which we showed
that clusters with $Q=0.3$ and $D=1.6$ (i.e., very cool and extremely
clumpy) undergo collapse to a short-lived but extremely dense core.
This core can dynamically mass segregate the most massive stars
down to a few M$_\odot$ via two-body encounters before re-expanding.
The re-expansion is partially driven by the increase in the velocity
dispersion of the low-mass stars caused by two-body encounters.

Our main results may be summarised as follows:

$\bullet$  The depth of the collapse, and so the degree of mass
segregation depend on both the initial virial ratio, $Q$, and the 
degree of substructure, $D$.  Low-$Q$ and low-$D$ clusters can
collapse to a denser state and so mass segregate more than high-$Q$,
high-$D$ clusters.

$\bullet$ Whilst there is a general trend of increasing mass
segregation with lower-$Q$ and lower-$D$ the inherently stochastic
nature of fractals means that {\em statistically} identical clusters may
undergo very different evolution.

$\bullet$  Many features are extremely short-lived, and young clusters
can change rapidly and violently.  Observations only provide a
snapshot of the evolution of a cluster and it is dangerous to draw
conclusions about the past and future state of a cluster from a single
snapshot (see also Bastian et al. 2008).

$\bullet$ Young clusters can undergo core collapse.  Early
dynamical mass segregation establishes a core of massive stars (see
also Pflamm-Altenburg \& Kroupa 2006) and a halo of lower-mass stars.
Energy loss from the core to the halo can drive the formation of
massive, hard binaries and Trapezium-like multiple systems.  The
heating of the halo and the hardening and subsequent decay of the
central multiple systems can dynamically destroy a cluster within a
few~Myr.

$\bullet$  The interactions of massive stars in the core of a young
cluster can cause the ejection of even very massive stars at
velocities in excess of $20$~km~s$^{-1}$.  This may help explain
ejections from the ONC (see also Pflamm-Altenburg \& Kroupa 2006).

$\bullet$ The early evolution of cool, clumpy clusters is rapid,
violent, and extreme.  The densities of clusters (and hence their
crossing and relaxation times) can change by orders of magnitude
during the first few~Myr of their existence.  Thus the currently
observed properties of young clusters are just a snapshot in the life
of these clusters and extreme care must be taken in inferring the past
history or future evolution of clusters from a single snapshot (see
also Goodwin \& Bastian 2006; Bastian et al. 2008; Allison et
al. 2009b).

That young clusters are mass segregated down to a few M$_\odot$, but
not below, is due to the short-lived dynamical mass segregation phase
which is able to mass segregate only the most massive stars.  That the
ONC has an unstable high-order multiple containing four of the most
massive stars can be explained by its dynamical formation during the
dense phase.  The ejection of high-mass stars from clusters can occur
during the dense phase, or afterwards from the decay of higher-order
massive multiple systems.

\section*{Acknowledgements}

RJA and RJP acknowledge financial support from STFC. SPZ is grateful
for the support of the Netherlands Advanced School in Astrophysics
(NOVA), the LKBF and the Netherlands Organisation for Scientific
Research (NWO) (via grants \#643.200.503 and \#639.073.803). RdG
acknowledges financial support from the National Science Foundation of
China under grant 11043006. We acknowledge the support and hospitality
of the International Space Science Institute in Bern, Switzerland
where part of this work was done as part of a International Team
Programme. We would also like to thank the referee, Ralf Klessen, for
his interesting and useful comments.

%-----------------------------------------------------------------------
%\newcommand{\bibfont}{\small}
%\setlength{\bibsep}{0pt}
%\bibliography{paper3bib}
\bibliographystyle{mn2e}

\end{document}